\documentclass{aa}
\usepackage{graphicx}
\usepackage{txfonts}
\usepackage{natbib}

\newcommand{\kms}{km~s$^{-1}$}
\newcommand{\ms}{m~s$^{-1}$}

\begin{document}

\title{Temporal evolution of the Evershed flow in sunspots\thanks{Appendices are
    only available in electronic form at \newline {\tt http://www.edpsciences.org}}}
\subtitle{I. Observational characterization of Evershed clouds}

\author{D.\ Cabrera Solana\inst{1}, L.R.\ Bellot Rubio\inst{1}, C.\ 
Beck\inst{2}, and J.C.\ del Toro Iniesta\inst{1} } 
\institute{Instituto de Astrof\'{\i}sica de Andaluc\'{\i}a, CSIC, Apdo.\ 
3004, 18080 Granada, Spain
\and Instituto de Astrof\'{\i}sica de Canarias, C/Via L\'actea s/n, 38200 La 
Laguna, Tenerife, Spain}

\offprints{L.R.\ Bellot Rubio}
\date{Received 8 May 2007 / Accepted }

\abstract {The magnetic and kinematic properties of the photospheric Evershed
flow are relatively well known, but we are still far from a complete
understanding of its nature. The evolution of the flow with time, which is
mainly due to appearance of velocity packets called Evershed clouds (ECs), may
provide information to further constrain its origin. }
{We undertake a detailed analysis of the evolution of the Evershed flow by
studying the properties of ECs. In this first paper we determine the sizes, 
proper motions, location in the penumbra, and frequency of appearance of ECs, 
as well as their typical Doppler velocities, linear and circular polarization 
signals, Stokes $V$ area asymmetries, and continuum intensities.}
{High-cadence, high-resolution, full vector spectropolarimetric measurements 
in visible and infrared lines are used to characterize the ECs phenomenon through
a simple line-parameter analysis as well as more complex Stokes inversions.}
{We find that ECs appear in the mid penumbra and propage outward
along filaments with large linear polarization signals and enhanced Evershed
flows.  The frequency of appearance of ECs varies between 15 and 40 minutes in
different filaments. ECs exhibit the largest Doppler velocities and
linear-to-circular polarization ratios of the whole penumbra.  In addition,
lines formed deeper in the atmosphere show larger Doppler velocities, much in
the same way as the ''quiescent'' Evershed flow. According to our
observations, ECs can be classified in two groups: type I ECs, which vanish in
the outer penumbra, and type II ECs, which cross the outer penumbral boundary
and enter the sunspot moat. Most of the observed ECs belong to type I. On
average, type II ECs can be detected as velocity structures outside of the
spot for only about 14 min. Their proper motions in the moat are significantly
reduced with respect to the ones they had in the penumbra.}
{}


\titlerunning{Observational characterization of Evershed clouds}
\authorrunning{Cabrera Solana et al.}
\maketitle

\section{Introduction}
The Evershed flow has been studied intensively over the last decades
\citep[for a review see, eg.][]{2003A&ARv..11..153S}. However, most of what we
know today about this dynamical phenomenon has been learned from observations
that do not reflect its temporal variability. The study of the evolution
of the flow requires stable observing conditions during long periods of time,
which is difficult without the help of adaptive optics systems. For this
reason, there are very few investigations published about the topic
\citep{1994ApJ...430..413S, 1994A&A...290..972R, 2003A&A...397..757R,
2003ApJ...584..509G}. Their main result is that the Evershed flow displays
variations on time scales of 8-25 min. This quasi-periodic behavior is
produced by the so-called Evershed clouds (ECs), velocity packets which
propagate from the mid penumbra to the outer penumbral boundary.

A good knowledge of ECs is crucial for a complete understanding of the nature
of the Evershed flow. Unfortunately, their observational properties are not
well characterized yet. Previous investigations were based on high resolution
filtergrams and Dopplergrams, but the polarization signals of ECs have never
been observed. As a result, important details such as the relation between
magnetic fields and ECs remain unknown.

In a series of two papers we present the first study on the temporal evolution
of the Evershed flow that makes use of high-cadence, high-precision, full
vector spectropolarimetric observations of visible and near-infrared
lines. The spatial resolution of these measurements, around 0\farcs6 at
1565~nm and 0\farcs7 at 630~nm, is one of the highest ever achieved in solar
spectropolarimetry.

This first paper concentrates on the observational characterization of ECs.
In Sect.~\ref{observations} we describe the observations and data reduction.
We then identify the ECs and determine their lifetimes, proper motions, and
sizes (Sect.~\ref{sec:identification}). Sections~\ref{sec:filaments} and
\ref{sec:periodicity} investigate the properties of the penumbral filaments
along which ECs move, and the periodicity of the EC phenomenon. We find that
ECs occur in the center of intra-spines, i.e., elongated structures having the
more horizontal fields and stronger Evershed flows of the penumbra
\citep{1993ApJ...418..928L}. The spectroscopic and polarimetric properties of
the ECs are presented in Sect.~\ref{sec:obs_prop}, and compared with those of
the intra-spines in Sect.~\ref{sec:comp_obs_prop}.  The dependence of the
Doppler velocities on height is examined in Sect.~\ref{sec:vlos_height}. 
We also study the correlation between proper motions and Doppler velocities 
(Sect.~\ref{sec:motions_vel_correla}), the evolution of the properties of the 
ECs as they propagate outward (Sect.~\ref{sec:rad_obs_prop}), and their 
disappearance inside (Sect.~\ref{sec:ec_dissapear}) and outside 
(Sect.~\ref{sec:out_pen_gen}) the penumbra. Our main findings are summarized in
Sect.~\ref{sec:conclusions_cap5}.

Paper II of the series \citep{cabrera.2007a} contains an analysis of the
thermal, magnetic, and dynamic properties of the ECs as inferred from
inversions of the measured Stokes spectra. Further details can be found in
\citet{cabrera.thesis}.

\section{Observations and data analysis}
\label{observations}
\subsection{Observations}

NOAA Active Region 10781 was observed on 2005 June 30 from 8:47 to 11:23 UT
and on 2005 July 1 from 9:31 to 10:51 UT at the German Vacuum Tower Telescope
(VTT) on Tenerife. The Tenerife Infrared Polarimeter
\citep[TIP;][]{1999ASPC..183..264M, 1999AGM....15..A13C} and the Polarimetric
Littrow Spectrograph \citep[POLIS;][]{2003AN....324..300S,
2005A&A...437.1159B} were operated simultaneously to record the full Stokes
profiles of the spectral lines around 1565~nm and 630~nm
(Table~\ref{tab:set_lines}). The same spot was observed on 2005 June 30 with
the Dutch Open Telescope (DOT) from 8:45 to 9:38~UT. Table~\ref{tab:obs_2005}
gives details of the observations at the VTT. The active region was located at
heliocentric angles of $43^\circ$ and $35^\circ$ on June 30 and July 1,
respectively. The slit width was 0\farcs36 for TIP and 0\farcs18 for
POLIS. During the observations, the Kiepenheuer Adaptive Optics System
\citep[KAOS;][]{2002AN....323..236S} was used to reduce image motion and
blurring.  In order to observe the same field of view (FOV) the TIP and POLIS
slits were aligned by rotating the main spectrograph of the VTT and moving the
TIP camera.  The pixel size was 0\farcs175 and 0\farcs 145 for TIP and POLIS,
respectively.

\begin{table}
\caption{Visible and infrared spectral lines observed with TIP and POLIS.
$\lambda_{0}$ represents the laboratory central wavelength, $\chi$ the
excitation potential of the lower level, and $\log \, gf$ the logarithm of 
the oscillator strength times the multiplicity of the lower level. $a$, $b$, 
and $c$ refer to \cite{1994ApJS...94..221N}, \cite{2003A&A...404..749B}, and
\cite{1974kptp.book.....P}, respectively. Note that ${\rm O_2}$ are telluric
absorption lines.
\label{tab:set_lines}}
\tabcolsep 0.3em
\begin{tabular}{l c c c c c } 
\hline
\hline
\multicolumn{1}{l}{Instrument} & \multicolumn{1}{c}{Species} & \multicolumn{1}{c}{$\lambda_{0}$} & 
\multicolumn{1}{c}{$\chi$} & 
\multicolumn{1}{c}{$\log{ gf}$} & \multicolumn{1}{c}{transition} \\ 
 & & [nm] & [eV] & &  \\
\hline
POLIS & Fe {\sc I}   & $630.1501^{a}$  & $3.65^{a}$ & $-0.72$ & $^{5}P_{2}$--$^{5}D_{2}$ \\ 
POLIS & Fe {\sc I}   & $630.2494^{a}$  & $3.69^{a}$ & $-1.24$ & $^{5}P_{1}$--$^{5}D_{0}$ \\ 
POLIS & Fe {\sc I}   & $630.3460^{ }$  & 4.32       & $-2.55$ & $^{5}G_{6}$--$^{5}G_{5}$      \\ 
POLIS & Ti {\sc I}   & $630.3753^{ }$  & 1.45       & $-1.44$ & $^{3}F_{3}$--$^{3}G_{3}$      \\ 
POLIS & ${\rm O_2}$  & $630.2001^{c}$  &            &                                         \\ 
POLIS & ${\rm O_2}$  & $630.2761^{c}$  &            &                                         \\ 
TIP   & Fe {\sc I}   & $1564.7410^{a}$ & $6.33^{a}$ & $-0.95$ & $^{7}D_{1}$--$^{5}P_{2}$ \\
TIP   & Fe {\sc I}   & $1564.8515^{a}$ & $5.43^{a}$ & $-0.68^{b}$ & $^{7}D_{1}$--$^{7}D_{1}$ \\
TIP   & Fe {\sc I}   & $1565.2874^{a}$ & $6.25^{a}$ & $-0.05^{b}$ & $^{7}D_{5}$--$^{6}D_{4.5}4f[3.5]^{0}_{4}$ \\
\hline
\end{tabular}
\end{table}

\begin{table}
\caption{Log of the VTT observations. The second column gives the heliocentric
angle of the spot. The third and fourth columns show the scanning step and the
integration time per slit position. The duration of the sequences and the
seeing conditions are indicated in the last two columns.
\label{tab:obs_2005}}
\tabcolsep .7em
\begin{tabular}{l c c c c c} 
\hline
\hline
\multicolumn{1}{c}{Date}  &\multicolumn{1}{c}{$\theta$} & 
\multicolumn{1}{c}{step size} & \multicolumn{1}{c}{$t_{\rm i}$} & 
\multicolumn{1}{c}{duration} & \multicolumn{1}{c}{seeing} \\
  & [deg] & [arcsec] & [s] & [min] & [arcsec] \\
\hline
06/30/2005 &  43 & 0.2 & 10.5 & 156 & $\sim 0.6$--0.7\\
07/01/2005 &  35 & 0.2 & 5.5  & 80  & $\sim 0.8$--0.9\\
\hline
\end{tabular}
\end{table}

\begin{figure}
\begin{center}
\scalebox{0.49}{\includegraphics{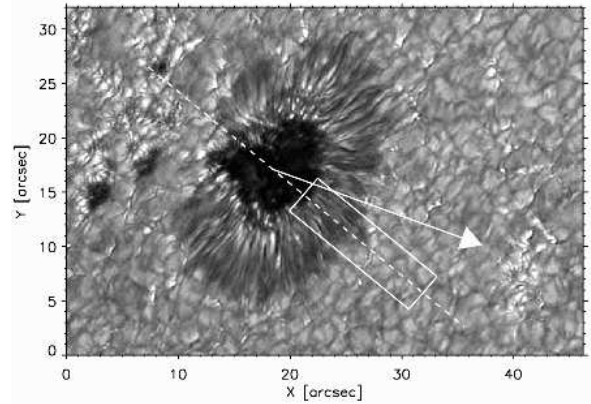}}
\caption{Speckle-reconstructed G-band filtergram of AR 10781 taken at
the DOT on 2005 June 30, 9:34 UT. The box shows
the fraction of the center-side penumbra and adjacent moat scanned by
TIP and POLIS. The arrow marks the direction to disk
center.\label{fig:time_seq_30jun_map}}
\end{center}
\end{figure}

On June 30, we performed rapid scans of a small portion of the center-side
penumbra of the spot including the adjacent moat region (see
Fig.~\ref{fig:time_seq_30jun_map}). The scan step was 0$\farcs$2 for a total
of 20 steps (4$\arcsec$). The integration time was 10.5~s per slit position,
resulting in a cadence of 3.9 min. We performed 40 repetitions of the
scan. The excellent seeing conditions, together with KAOS, allowed us to reach
a spatial resolution of about 0\farcs6 at 1565~nm and 0\farcs7 at
630~nm. These values have been derived from the power spectra of the quiet sun
granulation surrounding the spot.

On July 1 we observed a smaller portion of the center-side penumbra of the
same active region: the scan step was 0$\farcs$2 for a total of 10 steps,
producing maps of 2$\arcsec$. The integration time was 5.5~s per slit position
(2~min cadence). The seeing conditions were good and the spatial resolution
was around 0\farcs8 in the infrared and 0\farcs9 in the visible.

The use of infrared (IR) and visible (VIS) spectral lines, with different heights 
of formation, allows us to obtain physical information from a wide range of layers
in the atmosphere. Moreover, the different sensitivities of the spectral lines
to the atmospheric parameters improve the diagnostic capabilities of a single
spectral range alone \citep{2005A&A...439..687C}.

\subsection{Data reduction}

\subsubsection{Calibration of the polarimetric data}
To remove the different response of the CCD pixels to homogeneous
illumination, gain tables calculated using flat field and dark current images
were applied. Then, the intensity signals measured by the polarimeters can be
converted to Stokes profiles using the demodulation matrices derived with the
help of the instrumental calibration unit of the VTT. Instrumental crosstalk
was removed applying the telescope model of \cite{2005A&A...443.1047B}, while
residual crosstalk due to the coelostat mirrors was corrected for using the
statistical method of \cite{2003SPIE.4843...55C} and \cite{2002A&A...381..668S} . 
The accuracy of the polarization profiles is estimated to be on the order of 
$10^{-3}$ in units of the continuum intensity \citep{2005A&A...443.1047B}.

\subsubsection{Scattered light correction}
\label{sec:data_scatterlight}

The presence of residual O$_2$ telluric lines in the polarization profiles
recorded by POLIS reveals the existence of spectrally undispersed scattered
light (Fig.~\ref{fig:telluric_polis}). If a constant fraction of the incident
light enters the camera without going through the grating, the intensity
measurements are contaminated by a certain amount of light artificially
produced by the instrument. This scattered light might be due to incomplete
shielding of the light path inside the compact POLIS shelter.

We correct the spectra for scattered light in the same way as
\citet{2004A&A...423.1109A}.  The amount of scattered light and the resolving
power of the polarimeters are calculated by comparing their average quiet-sun
intensity profiles with those in the Fourier Transform Spectrometer (FTS)
atlas. The FTS profiles are not affected by scattered light and the point
spread function (PSF) of the FTS can be considered as a Dirac delta function
for our purposes. Hence, if we add a certain fraction of scattered light to
the FTS profiles and convolve them with the spectral PSF of the polarimeters,
we should be able to reproduce the observed line shapes, i.e.,
\begin{equation}
I = \left(\frac{I_{\rm FTS}+K}{1+K}\right) \ast G(\sigma),
\label{eq:scatter_light}
\end{equation}
where $I$ are the observed disk-center profiles, $I_{\rm FTS}$ the FTS
profiles, $K$ is the fraction of scattered light, $G$ the polarimeter
instrumental profile (assumed to be a Gaussian of width $\sigma$), and $\ast$
represents convolution.

Figure~\ref{fig:scatter_polis} displays an example of the fits obtained using
Eq.~\ref{eq:scatter_light}. The best results are achieved with $K \simeq 15 \%$
and $\sigma \simeq 6$~pm, both for the July 1 and June 30 data sets. The
theoretical resolving power of POLIS (2~pm, limited by the slit) is always
smaller than that inferred from the fits. The somewhat large value of 
$\sigma$ may be a consequence of imperfect focussing of the instrument.

\subsection{Wavelength scale}
\label{sec:wave_scale}
The velocity scale for the IR and VIS lines has been set assuming that the
umbra does not move \citep{1977ApJ...213..900B}. The wavelength scale is
defined by the rest wavelength and the spectral dispersion. We obtain the
former calculating the mean zero-crossing positions of Stokes $V$ for the
pixels located in the darkest regions of the umbra\footnote{For the POLIS
data, the small focal length of the spectrograph produce a curvature of the
spectrum along the slit. In order to remove it, we use the O$_2$ telluric line
at $630.20$~nm to align all the spectra individually. This correction has been
applied before the determination of the wavelength scale.}. For the June 30
time sequence, we select pixels with continuum intensities $I/I_{\rm qs}<0.4$
at 630~nm. The dispersion is computed by determining the number of pixels
between the line core positions of \ion{Fe}{i} 1564.85~nm and \ion{Fe}{i}
1565.52~nm for TIP and the ${\rm O_{2}}$ telluric lines for POLIS. 

To estimate a lower limit for the calibration error, we calculate the
dispersion of the Doppler shifts in the selected umbral pixels with respect to
their mean value for each map. In Table~\ref{tab:cal_vel} we show the average
of those quantities in the two sequences. The errors indicate the standard
deviations of the averages. As can be seen, the mean dispersions are always
smaller than $120$~\ms~with low scatter (rms $<34$~\ms) between maps. 
Similar dispersions have been found by \cite{2006A&A...454..975R} in their
analysis of velocities in sunspot umbrae. Such small values could be produced
by the non-zero velocities of umbral dots \citep{2004ApJ...614..448S} and/or
the presence of oscillations in the selected areas.

\begin{table}
\caption{Mean dispersions of the Doppler shifts of the selected umbral pixels
for the different spectral lines. The errors represent the standard deviations
of the averages.
\label{tab:cal_vel}}
\tabcolsep .7em
\begin{tabular}{l c c c c c} 
\hline \hline \multicolumn{1}{c}{Date} & \multicolumn{1}{c}{$\sigma_{630.15}$}
&\multicolumn{1}{c}{$\sigma_{630.25}$} & \multicolumn{1}{c}{$\sigma_{630.35}$}
& \multicolumn{1}{c}{$\sigma_{1564.85}$} &
\multicolumn{1}{c}{$\sigma_{1565.29}$} \\ & [\ms] & [\ms] & [\ms] & [\ms] &
[\ms] \\ \hline 06/30 & $83\pm16$ & $80\pm17$ & $100\pm20$ & $86\pm16$ &
$120\pm16$ \\ 07/01 & $64\pm30$ & $63\pm31$ & $74\pm34$ & $64\pm28$ &
$60\pm28$ \\ \hline
\end{tabular}
\end{table}

\begin{figure}
\centering
\scalebox{0.7}{\includegraphics{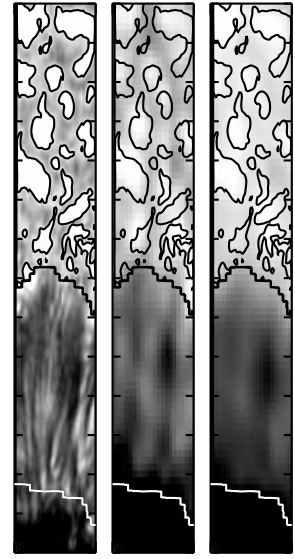}}
\caption{From left to right: Maps of speckle-reconstructed G-band intensity,
continuum intensity at 1565~nm, and continuum intensity at 630~nm
observed on June 30, 9:26~UT. Contours outline pixels with G-band
intensities greater than $0.5 \, I_{\rm qs}$. Each tickmark on the
$y$-axis represents $1\arcsec$.\label{fig:aligned}}
\end{figure}

\subsection{Alignment of TIP, POLIS, and DOT data}
\label{sec:alignment}

\subsubsection{Alignment of TIP and POLIS}

Since the refractive index of the air depends on wavelength, the images of the
Sun taken by TIP and POLIS are shifted with respect to each other
\citep{2006SoPh..239..503R}. In order to remove these shifts, we follow the
method described by \cite{2007.beck_a}.  It basically consists in performing a
cross-correlation of the intensity maps using the largest pixel (the TIP
pixel) as a reference. To that aim, the POLIS data are rebinned to the TIP
pixel size.  Figure~\ref{fig:aligned} displays, as an example, continuum
intensity maps at 1565~nm and 630~nm after removing the shifts. As can be
seen, the correspondence between the intensity structures is quite
precise. The accuracy of the alignment between TIP and POLIS is around $0
\farcs 1$ \citep{2007.beck_a}.

The position of the Sun on the sky changes during the day. Hence, the
differential refraction produces time-dependent shifts between the VIS and IR
spectropolarimetric data. When the time sequences are long, one has to take
into account the temporal dependence of the displacements.  This is why we
calculate the displacements between IR and VIS spectra for each repetition of
the scan. The total displacements accumulated during the 156~min of
observations on June 30 are around $1\arcsec$ and $0 \farcs 6$ in the
directions perpendicular and parallel to the slit, respectively. This means
that co-spatial Stokes profiles in the visible and the near-infrared are not
strictly simultaneous, but the time difference is always smaller than $\sim
55$~s.

\subsubsection{Alignment of VTT and DOT data}

\begin{figure*}
\begin{center}
\scalebox{0.712}{\includegraphics[bb=60 435 790 1005,clip]{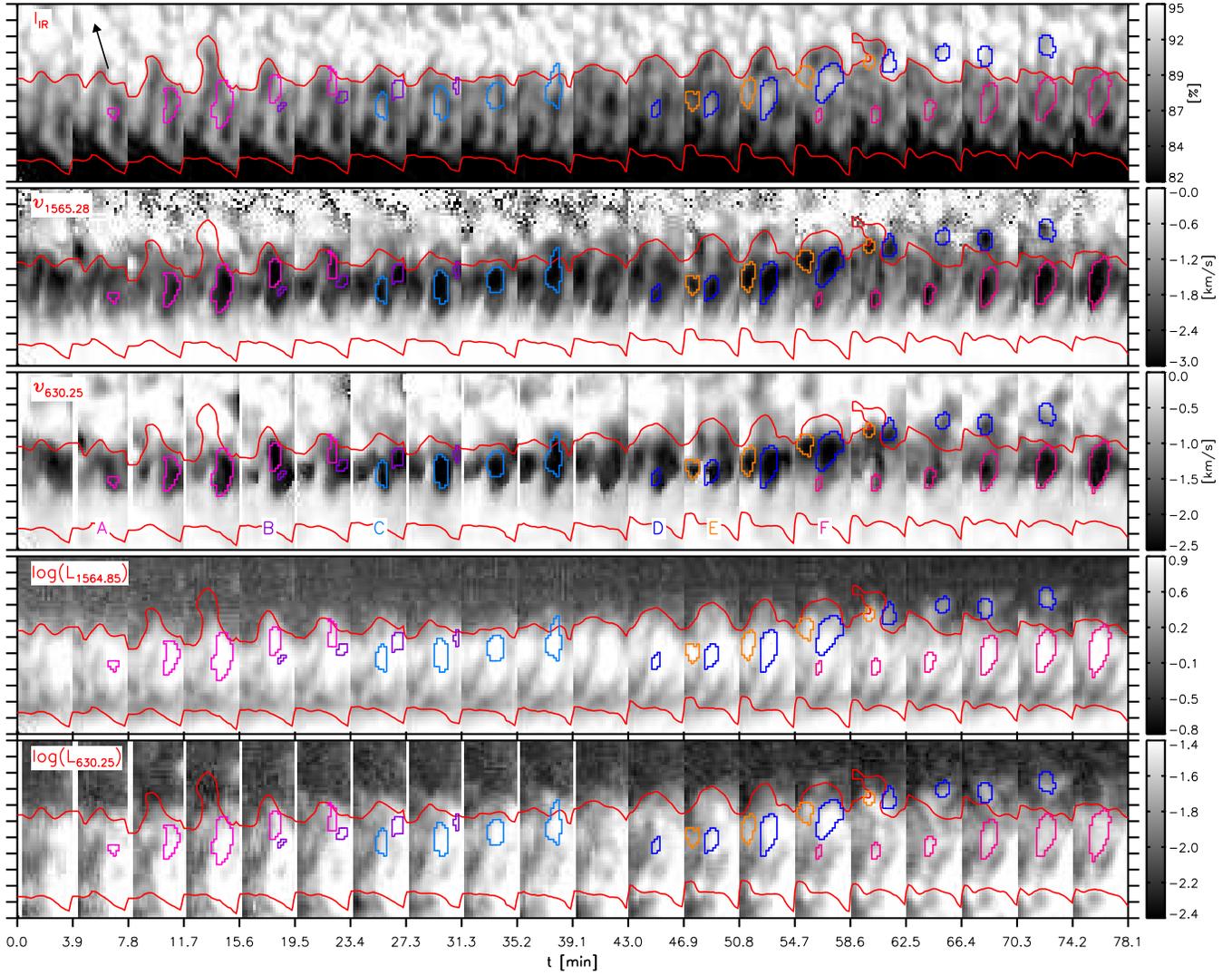}}
\caption{Evolution of line parameters in the center-side penumbra of AR
10781 on June 30. {\em Top} to {\em bottom}: Continuum intensity at
1565~nm, Stokes $V$ zero-crossing velocity of \ion{Fe}{i} 1565.28~nm and
\ion{Fe}{i} 630.25~nm, logarithm of the total linear polarization of
\ion{Fe}{i} 1564.85~nm and \ion{Fe}{i} 630.25~nm. Color contours outline the
ECs. The letters at the bottom of the third panel label each EC. Red lines
indicate the inner and outer penumbral boundaries. Each tickmark in the
$y$-axis represent 1\arcsec. The arrow marks the direction to disk
center. $t=0$ min corresponds to June 30, 8:47 UT.\label{fig:30general_1}}
\end{center}
\end{figure*}

\begin{figure*}
\begin{center}
\scalebox{0.712}{\includegraphics[bb=60 435 790 1005,clip]{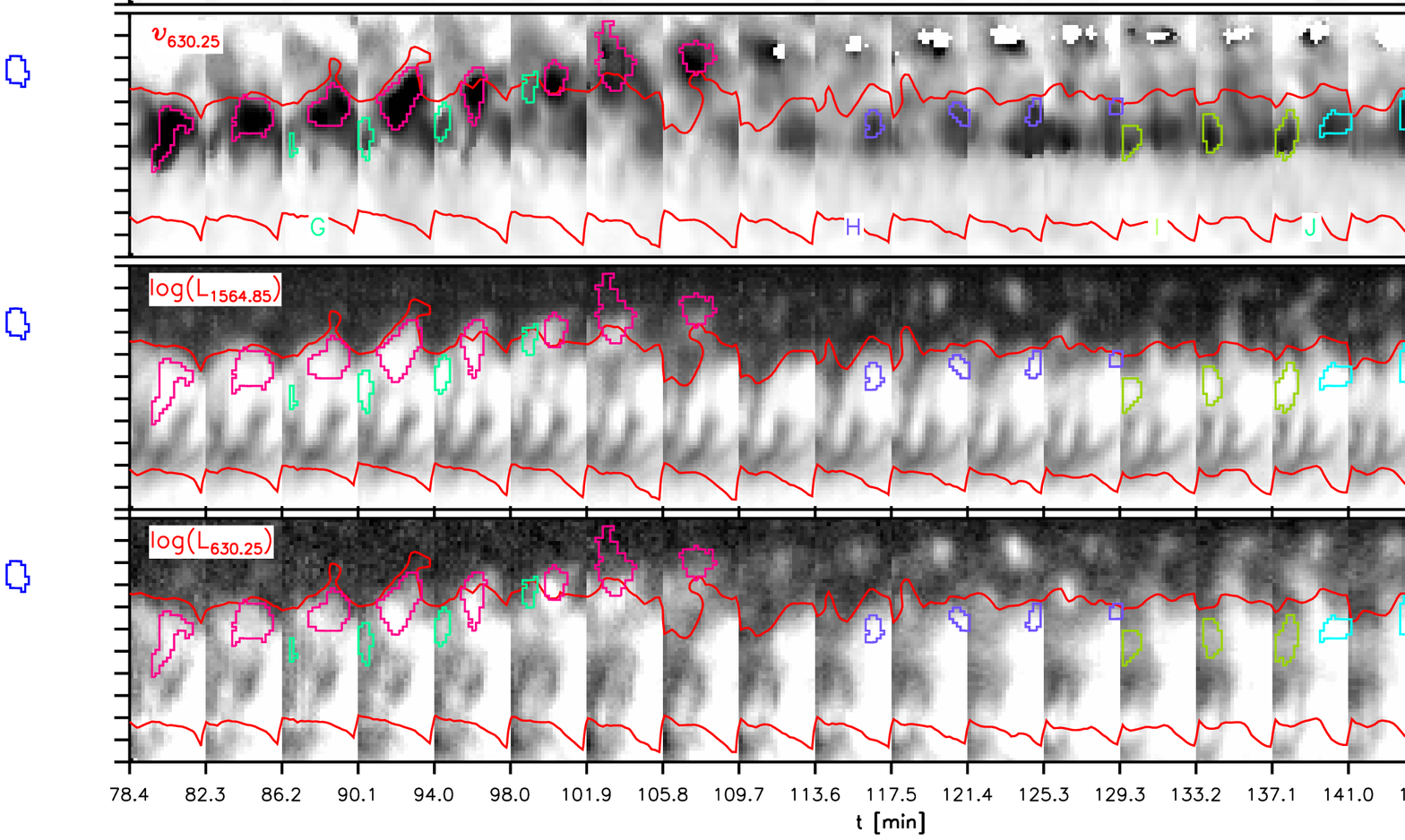}}
\caption{Same as Fig.~\ref{fig:30general_1}, from $t=78.4$ to
$t=156.6$ min.\label{fig:30general_2}}
\end{center}
\end{figure*}

The way filtergrams and spectropolarimetric maps are obtained is
different. The DOT filtergrams are 2D images taken at once, while the maps
recorded at the VTT are constructed by scanning the slit across the solar
surface, which takes time. If one wants to compare both kind of observations,
the positions observed by the polarimeters have to be located on the DOT
filtergrams taken at the same time. Following \cite{2007.beck_a}, we align the
POLIS slit-jaw images with the DOT G-band images. This is done by degrading
the G-band data to the pixel size of the slit-jaw images.  After the rotation
between the images of the two telescopes is removed, shifts between the
slit-jaw and the G-band images are determined by cross-correlation. At this
point, one knows where the positions of the POLIS/TIP slits are in the G-band
images. Finally, one constructs {\em artificial} G-band maps by selecting the
co-spatial G-band strip which is closer in time to the polarimetric data. The
DOT and VTT observations are not strictly co-temporal, but the differences are
small enough ($\le 30$~s) to work with both data sets as if they had been
taken simultaneously.

\subsection{Line parameters}
The simplest way to analyze the observations is by computing line parameters
from the polarization profiles. Those parameters reflect the atmospheric
conditions of the solar photosphere, providing a first estimate of their
values. We have derived the following observables from the Stokes profiles 
of the IR and VIS spectral lines:

\begin{enumerate}
\item Continuum intensities, $I$. According to the Eddintong-Barbier
approximation, $I$ depends on the temperature of the layers where the
continuum is formed [$\sim \log{(\tau)} = 0$]. This approximation works
well in sunspot penumbrae \citep{1994ApJ...436..400D,2006A&A...453.1117B}.

\item Stokes $V$ zero-crossing velocity. The Doppler velocity is estimated 
from the Doppler shift of the Stokes-$V$ zero-crossing wavelength,
$\lambda_{ZC}$. $\lambda_{ZC}$ is calculated as the mid-point between the 
maxima of the Stokes $V$ lobes. This velocity is representative of the 
magnetic atmosphere inside the resolution element and can only be 
computed for pixels exhibiting normal two-lobed Stokes $V$ profiles.

\item Total polarization signals. For fully split lines, if the
magnetic field orientation remains constant along the LOS
\citep{1993SSRv...63....1S}:

\begin{itemize}

\item The total linear polarization, $L = \sum_{i} \sqrt{Q(\lambda_{i})^2 +
  U(\lambda_{i})^2}$, is proportional to $\sin^2{\gamma_{\rm LOS}}$. Hence,
  the larger the value of $L$, the greater the inclination of the magnetic 
  field to the LOS ($\gamma_{\rm LOS}$).

\item The total circular polarization, $V = \sum_{i} |V(\lambda_{i})|$, is
  proportional to $\cos{\gamma_{\rm LOS}}$. Therefore it is smaller when the
  field inclination is larger.

\item The linear-to-circular polarization ratio is greater when the
inclination increases ($L/V \sim \sin^2{\gamma_{\rm LOS}}/\cos{\gamma_{\rm
LOS}}$).

\end{itemize}

The polarization signals provide only a rough estimate of $\gamma_{\rm LOS}$,
because the magnetic field is known to vary with height in the solar
photosphere and the spectral lines are not always in the strong field regime.

\item The Stokes $V$ area asymmetry, given by

\begin{equation}
\delta A = \frac{|A_{\rm r}|-|A_{\rm b}|}{|A_{\rm r}|+|A_{\rm b}|},
\end{equation}
where $A_{\rm r}$ and $A_{\rm b}$ are the areas of the red and blue lobes of
Stokes $V$.  $\delta A$ arises as a consequence of velocity gradients along
the LOS \citep{1978A&A....64...67A,1992ApJ...398..359S}. The combination of
such gradients with magnetic field or temperature gradients leads to an
amplification of the asymmetries.

\item We construct magnetograms, $M$, as (minus) the Stokes $V$ signal of
\ion{Fe}{i} 630.25 nm at $\Delta \lambda=+10$ pm from line center. $M$ 
provides a rough estimate of the longitudinal magnetic flux 
\citep{1992soti.book...71D}.

\end{enumerate}

\section{Identification, morphology, and proper motions of ECs}
\label{sec:identification}


\begin{table*}[t]
\tabcolsep .77em
\caption{Phenomenological properties
of the observed ECs. ECs reaching the quiet photosphere are marked 
in bold face. Parentheses indicate ECs that leave or appear outside of the
FOV. The date and time of appearance of each EC are specified in the
second and third columns. The fourth column displays the radial
distance at which the EC is seen for the first time. Only the values
for the ECs appearing inside the FOV are given. The fifth column
displays the time each EC stays inside the penumbra. The seventh
column shows the propagation velocity, corrected for the viewing
angle. The errors are the standard deviations in the determination of
the proper motions.  The last two columns display the maximum lengths
and widths attained by each EC, corrected for projections effects.
\label{tab:ecs_prop1}}
\begin{tabular}{l c r c r c c r r}
\hline
\hline
\multicolumn{1}{l}{EC} & \multicolumn{1}{c}{Date} &
\multicolumn{1}{c}{$t_{\rm app}$} & \multicolumn{1}{c}{$r_{\rm app}$}
& \multicolumn{1}{c}{$\tau$} & \multicolumn{1}{c}{Filament} &
\multicolumn{1}{c}{$v_{\rm prop}$} & \multicolumn{1}{c}{Length} &
\multicolumn{1}{c}{Width}\\ & & [min] & & [min] & & [\kms] & \multicolumn{1}{c}{[km]} & \multicolumn{1}{c}{[km]} \\
\hline

A       & 06/30  & 3.9   & 0.4 & 15.6   & 2 & $2.7 \pm 0.5$ & 2700 & 1000    \\ 
(B)     & 06/30  & 15.6  & 0.5 &        & 2 & $2.9 \pm 0.0$ & 500  & 400     \\ 
C       & 06/30  & 23.4  & 0.4 & 11.7   & 2 & $2.2 \pm 0.7$ & 2600 & 800     \\ 
{\bf D} & 06/30  & 42.9  & 0.5 & 11.9   & 2 & $3.3 \pm 0.3$ & 2300 & 900     \\ 
E       & 06/30  & 46.8  & 0.5 & 11.9   & 1 & $3.1 \pm 0.4$ & 1700 & 800     \\ 
{\bf F} & 06/30  & 54.8  & 0.4 & 39.0   & 2 & $1.7 \pm 0.1$ & 3000 & 1200    \\ 
G       & 06/30  & 86.0  & 0.6 & 11.7   & 1 & $3.7 \pm 0.8$ & 1700 & 500     \\ 
(H)     & 06/30  & 113.5 & 0.7 &        & 3 & $1.7 \pm 0.0$ & 1000 & 500     \\ 
(I)     & 06/30  & 129.1 &     &        & 1 & $1.5 \pm 0.4$ & 2000 & 800     \\ 
J       & 06/30  & 136.9 & 0.7 & 3.9    & 3 & $3.5 \pm 0.0$ & 1500 & 800     \\ 
K       & 06/30  & 144.7 & 0.5 & 7.8    & 1 & $2.1 \pm 0.6$ & 1700 & 800     \\ 
(L)     & 07/01  & 5.9   & 0.7 &        &   & $2.2 \pm 0.8$ & 1200 & 600     \\ 
{\bf M} & 07/01  & 9.9   & 0.7 & 7.9    &   & $4.2 \pm 0.6$ & 1800 & 800     \\ 
({\bf N})& 07/01 & 23.8  &     &        &   & $2.0 \pm 0.2$ & 1200 & 500     \\ 
(O)      & 07/01 & 61.6  & 0.8 &        &   & $1.4 \pm 0.5$ & 1100 & 600     \\
\hline
{\bf Mean} &  &   & {\bf 0.6} & {\bf 13.5}&   &   {\bf 2.6}     &{\bf 1700}&{\bf 700}     \\
\hline
\end{tabular}
\end{table*}

In Figs.~\ref{fig:30general_1} and \ref{fig:30general_2} we show the temporal
evolution of the center-side penumbra and the moat region of AR 10781 during
the 156 min of the June 30 data set.  The first panel displays the continuum
intensity at 1565~nm. The next two panels display the Doppler velocities
calculated from the Stokes $V$ zero-crossing shifts of \ion{Fe}{i} 1565.28~nm
and \ion{Fe}{i} 630.25 nm\footnote{The velocity is computed only for pixels
exhibiting normal two-lobed $V$ profiles with amplitudes greater than 0.2$\%$
for \ion{Fe}{i} 630.25~nm and 0.5$\%$ for \ion{Fe}{i} 1564.85~nm.}. The last
two panels show the logarithm of the total linear polarization of \ion{Fe}{i}
1564.85~nm and \ion{Fe}{i} 630.25~nm.  Appendix \ref{sec:appendix} gives the
same information for the July 1 data set.

We identify ECs as structures of enhanced velocity signal which propagate
outward across the penumbra. During the 236 minutes of our observations,
fifteen ECs appear in the mid-penumbra and then migrate to the outer penumbral
boundary. They are observed in the two spectral ranges simultaneously. We have
outlined them with contours of different colors and letters in
Figs.~\ref{fig:30general_1} and \ref{fig:30general_2}. The boundaries of the
ECs have been defined as isocontours of specific values of the Doppler
velocity, hence they should be considered as approximate.

In order to quantify the radial distances where the ECs appear, we calculate
the temporal average of the penumbral borders and define the mean penumbral
radial distance, $r$, as the normalized distance from the inner ($r = 0$) to
the outer ($r = 1$) mean edges of the penumbra.

Table~\ref{tab:ecs_prop1} summarizes some phenomenological properties of the
observed ECs. The radial distances where they appear, $r_{\rm app}$, vary from
0.4 to 0.7 with an average of 0.6.  Their lifetimes, $\tau$, inside the
penumbra range from 3.9 to 39~min with an average of 13.5~min\footnote{$\tau$
is only calculated for the ECs that do not leave the FOV}. ECs can be
classified in two different groups: (a) those that disappear at the outer
penumbral boundary, hereafter type {\rm I} ECs; and (b) those that reach the
quiet photosphere and enter the moat (D, F, M, and N), type {\rm II} in the
following. Most of the ECs belong to the first group.

The propagation velocities of the ECs inside the penumbra, corrected for the
viewing angle, range from 1.4~km~s$^{-1}$ to 4.2~km~s$^{-1}$ with an average
of 2.6~km~s$^{-1}$ (seventh column of Table~\ref{tab:ecs_prop1}).  These
velocities are in rough agreement with values reported earlier
\citep{1994ApJ...430..413S, 1994A&A...290..972R, 2003ApJ...584..509G}.  To
compute the propagation velocity of each EC we have fitted straight lines to
the ($x,y$) positions of the edge farther from the umbra as a function of
time. The slopes of these lines give the apparent propagation velocities along
the line of symmetry\footnote{The line of symmetry is the projection of the
LOS onto the solar surface. In the case of sunspots, it coincides with the
line connecting the sunspot center and the disk center.} (LS) and
perpendicularly to it.  Due to projection effects the distances are contracted
along the LS. We have corrected for this assuming that ECs move parallel to
the solar surface. Note that the vertical motion of ECs must be very small,
otherwise they would leave the line forming region in a few
minutes\footnote{The thickness of the photosphere is considered to be around
500~km. If the vertical motions were greater than 600~m/s, ECs departing from
continuum forming layers would cross the entire photosphere in less than their
mean lifetime ($\sim 13.5$~min).}.

\begin{figure}[h]
\begin{center}
\scalebox{0.37}{\includegraphics{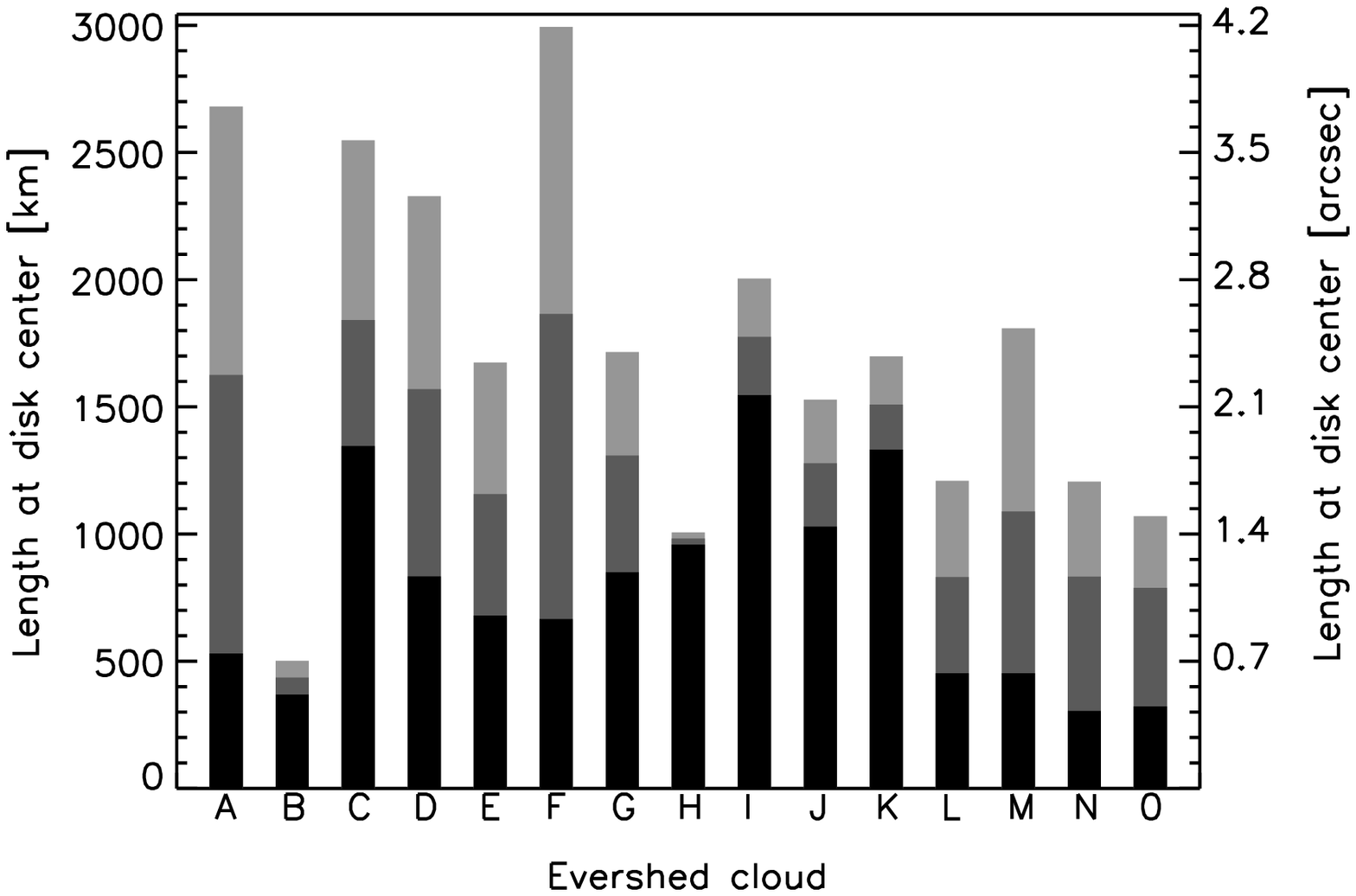}}
\scalebox{0.37}{\includegraphics{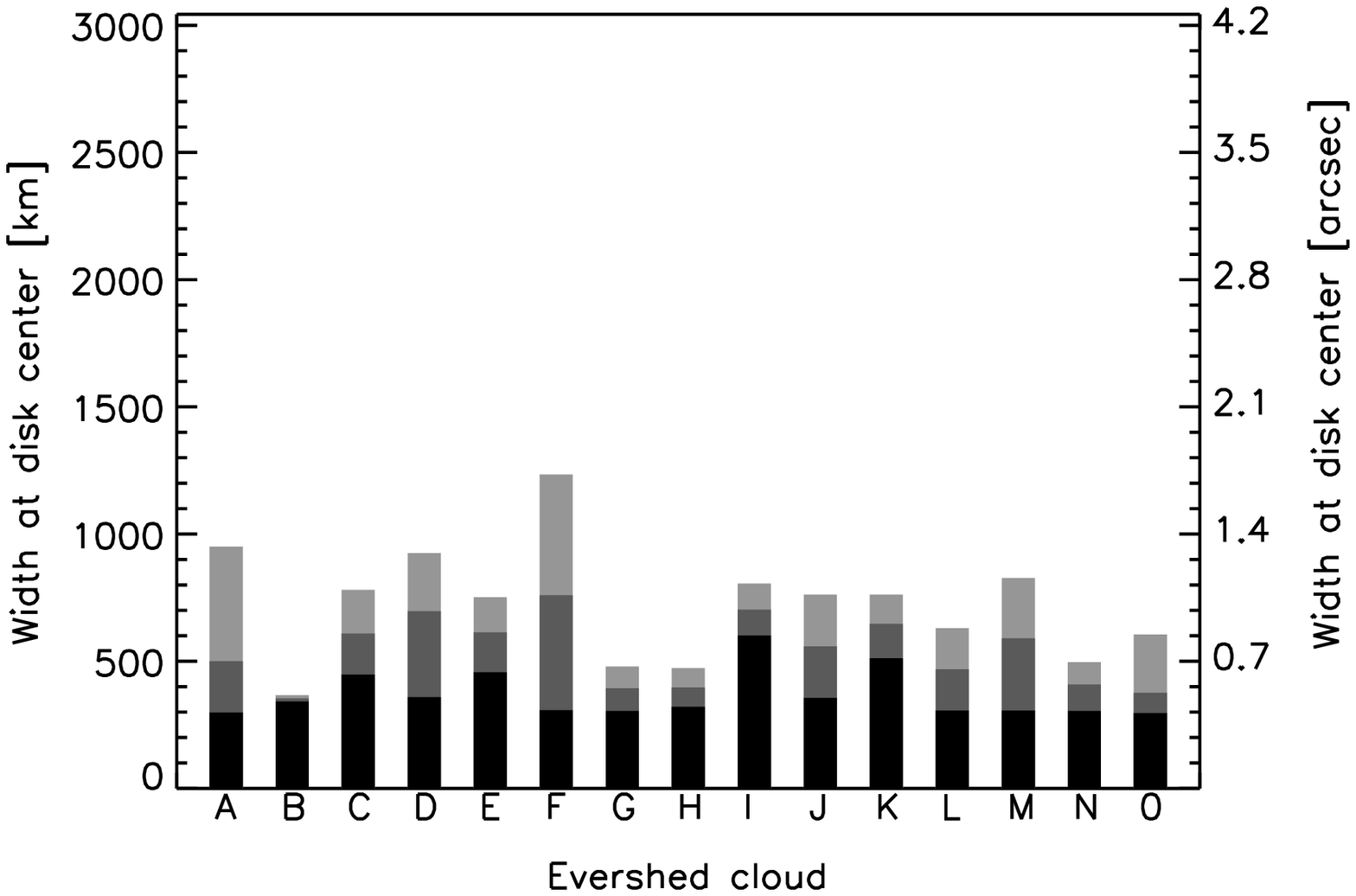}}
\caption{Maximum, mean, and minimum lengths ({\em top}) 
and widths ({\em bottom}) of the ECs, corrected for
projection effects (light to dark gray). \label{fig:sizes}\vspace*{-2em}}
\end{center}
\end{figure}

\begin{figure*}
\begin{center}
\scalebox{0.41}{\includegraphics{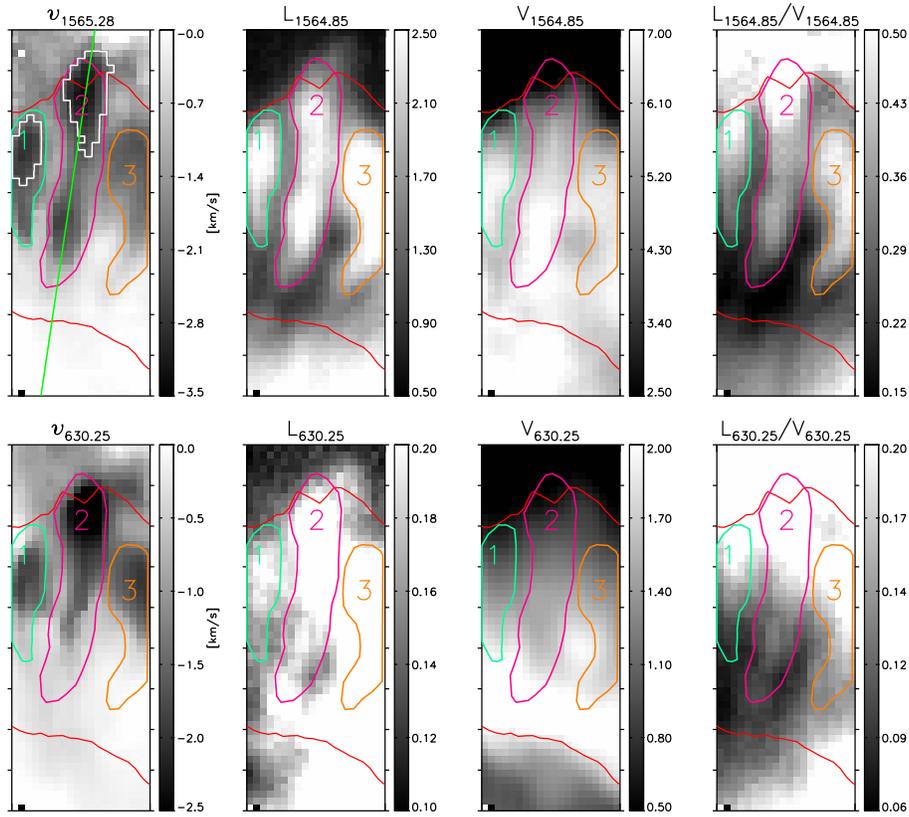}}
\caption{From {\em left} to {\em right}: Maps of Stokes $V$ zero-crossing
velocity for \ion{Fe}{i} 1565.28~nm (top) and \ion{Fe}{i} 630.25~nm (bottom),
total linear polarization, total circular polarization, and $L/V$ ratio for
\ion{Fe}{i} 1564.85~nm (top) and \ion{Fe}{i} 630.25~nm (bottom) at $t=93.8$
min. Color contours mark, and numbers label, three different penumbral
filaments. White contours outline ECs F and G. The green line represents a cut
along filament \#2. \label{fig:filaments}}
\end{center}
\end{figure*}

In our observations, most ECs appear as elongated structures along the radial
direction, and their sizes usually increase as they migrate outwards. To
quantify the typical sizes of the ECs we define a major and a minor axis for
each one. The major axis is the segment connecting the edges of the EC along
its longer direction. The minor axis is perpendicular to the major axis and
passes through its midpoint. We compute the length and width of the ECs along
the two axes and correct them for projection effects. The maximum length of
the ECs varies from 500 to 3000~km with an average of 1700~km, while the
maximum width ranges from 400 to 1200~km with an average of 700~km (cf.\ the
two last columns of Table~\ref{tab:ecs_prop1}).  Note that for these
computations we only use the ECs which are not partially outside the FOV. In
Fig.~\ref{fig:sizes} we show histograms of the maximum, minimum, and mean
sizes along the major and minor axes for each EC. This figure also
demonstrates that the lengths are usually greater than the widths.

\cite{1994ApJ...430..413S} described the ECs as cloud-like structures with
size roughly 1000 km in length and a radial spacing of about 2000~km between
subsequent ECs. Our lengths are slightly larger than those of
\cite{1994ApJ...430..413S}, while the typical separation between two
consecutive ECs is around 500~km in our observations (cf.\ second panels of
Fig.~\ref{fig:30general_1} at $t=15.6$ and $t=54.8$~min).

\section{Relation with penumbral filaments}
\label{sec:filaments}

The second, third, and fourth panels of Figs.~\ref{fig:30general_1} and 
\ref{fig:30general_2} demonstrate that ECs move outward following penumbral 
filaments with larger Doppler velocities and total linear polarization signals 
than their surroundings. At least three different penumbral filaments can 
be distinguished in the June 30 data set. One of them is visible in the middle 
of the maps during the whole time sequence. The other two are partially outside 
of the FOV. We label the left-side, center, and right-side filaments as \#1, 
\#2, and \#3, respectively. The sixth column of Table~\ref{tab:ecs_prop1} 
identifies the filament along which each EC propagates outward.

\begin{figure}
\begin{center}
\scalebox{0.65}{\includegraphics{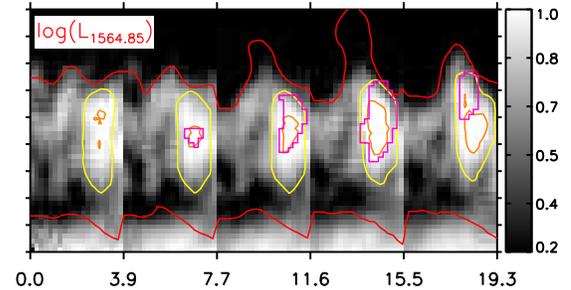}}
\caption{Logarithm of the total linear polarization of \ion{Fe}{i} 1564.85~nm
as a function of time. Yellow and pink contours mark filament $\#$2 and EC A,
respectively. Orange contours outline pixels with $\log{ L_{1564.85}}>1$.
\label{fig:ec_filament}}
\end{center}
\end{figure}

Figure~\ref{fig:filaments} blows up the velocity and polarization maps of the
penumbra at $t=93.8 $~min. The various panels display Doppler velocities,
total linear polarizations, and total circular polarizations, along with their
ratio, $L/V$. Penumbral filaments are outlined with contours of different colors
which have been constructed taking into account the velocity {\em and} linear
polarization signal maps. As can be seen, the filaments show up as structures
with enhanced $v$, $L$, and $L/V$, and their contrast is much higher in the
quantities derived from the infrared lines. They exhibit larger circular
polarization signals than their surroundings in \ion{Fe}{i} 1564.85~nm, but
smaller signals in \ion{Fe}{i}~630.25~nm. These properties indicate that the
filaments observed in the maps are penumbral intra-spines \citep{1993ApJ...418..928L}.

\begin{table*}[t]
\caption{Maximum Doppler shifts, linear-to-circular
polarization ratios, and Stokes $V$ asymmetries attained by each
EC. The second and third columns are the unsigned blueshifts of
\ion{Fe}{i} 1564.85~nm and \ion{Fe}{i} 630.25 nm. The fourth and fifth
columns display the ratios of total linear and circular polarization
of \ion{Fe}{i} 1564.85~nm and \ion{Fe}{i} 630.25~nm. The last two
columns are the area asymmetries of \ion{Fe}{i} 1564.85~nm and
\ion{Fe}{i} 630.25~nm. The errors represent the standard deviation 
of the individual parameters within each EC. The mean values of the
observational parameters for the June 30 and July 1 data sets are
marked in bold face.\label{tab:ecs_prop2}}
\begin{tabular}{l c c c c c r} 
\hline
\hline
\multicolumn{1}{l}{EC} & \multicolumn{1}{c}{$|v_{\rm 1564.85}|$} &
\multicolumn{1}{c}{$|v_{\rm 630.25}|$} &
\multicolumn{1}{c}{$(L/V)_{\rm 1564.85}$}
&\multicolumn{1}{c}{$(L/V)_{\rm 630.25}$} & \multicolumn{1}{c}{$\delta
A_{\rm 1564.85}$} & \multicolumn{1}{c}{$\delta A_{\rm 630.25}$}\\ &
[\kms] & [\kms] & & & [$\%$] & \multicolumn{1}{c}{[$\%$]} \\ \hline
A & $3.3\pm0.3$ & $2.4\pm0.2$ & $0.44\pm0.02$ & $0.19\pm0.03$ & $5.9\pm2.0$  & $2.4\pm2.3$     \\ 
B & $3.4\pm0.3$ & $2.5\pm0.3$ & $0.50\pm0.05$ & $0.27\pm0.02$ & $7.3\pm3.2$  & $7.7\pm1.8$     \\ 
C & $3.2\pm0.3$ & $2.6\pm0.3$ & $0.45\pm0.03$ & $0.20\pm0.03$ & $6.9\pm1.7$  & $4.2\pm3.3$     \\ 
D & $3.2\pm0.5$ & $2.4\pm0.5$ & $0.44\pm0.07$ & $0.22\pm0.04$ & $5.8\pm2.4$  & $5.1\pm2.8$     \\ 
E & $3.1\pm0.3$ & $2.1\pm0.4$ & $0.44\pm0.06$ & $0.27\pm0.02$ & $5.3\pm1.9$  & $7.0\pm3.7$     \\ 
F & $3.4\pm0.5$ & $2.5\pm0.4$ & $0.56\pm0.14$ & $0.29\pm0.05$ & $6.5\pm2.4$  & $9.6\pm6.5$     \\ 
G & $3.0\pm0.3$ & $1.5\pm0.1$ & $0.55\pm0.05$ & $0.27\pm0.03$ & $9.1\pm3.7$  & $10.1\pm1.6$     \\ 
H & $3.1\pm0.2$ & $1.9\pm0.3$ & $0.49\pm0.02$ & $0.24\pm0.01$ & $6.1\pm1.7$  & $5.8\pm2.5$     \\ 
I & $3.1\pm0.2$ & $1.9\pm0.2$ & $0.36\pm0.04$ & $0.14\pm0.02$ & $3.6\pm1.3$  & $-2.4\pm1.3$     \\ 
J & $3.2\pm0.3$ & $2.3\pm0.3$ & $0.49\pm0.02$ & $0.24\pm0.01$ & $5.5\pm2.6$  & $1.7\pm2.5$    \\ 
K & $3.1\pm0.2$ & $1.5\pm0.4$ & $0.41\pm0.05$ & $0.16\pm0.02$ & $3.2\pm1.5$  & $-3.2\pm2.3$     \\ 
\hline
Mean  & {\bf 3.2}  & {\bf 2.2}   &  {\bf 0.47}   & {\bf 0.22}    &  {\bf 5.9}   &  {\bf 4.4}  \\  \hline
L & $2.1\pm0.1$ & $1.4\pm0.1$ & $0.87\pm0.07$ & $0.37\pm0.03$ & $4.7\pm1.0$ & $-2.1\pm1.1$    \\ 
M & $2.2\pm0.1$ & $1.5\pm0.1$ & $0.83\pm0.06$ & $0.41\pm0.04$ & $5.5\pm1.4$ & $4.7\pm3.4$     \\
N & $2.1\pm0.2$ & $1.2\pm0.2$ & $0.83\pm0.08$ & $0.38\pm0.03$ & $5.9\pm2.5$ & $6.9\pm4.7$    \\ 
O & $1.9\pm0.3$ & $1.0\pm0.1$ & $0.69\pm0.02$ & $0.31\pm0.02$ & $5.4\pm2.7$ & $3.3\pm1.8$   \\ 
\hline
Mean  & {\bf 2.1}  & {\bf 1.3}   &  {\bf 0.81}   & {\bf 0.37}    &  {\bf 5.4}   &  {\bf 3.2}     \\ 
\hline
\end{tabular}
\end{table*}

In Fig.~\ref{fig:ec_filament} the total linear polarization of \ion{Fe}{i}
1564.85~nm is shown as a function of time for the maps where EC A
is detected. Pink and yellow contours outline EC A and filament $\#$2,
respectively. While the shape of the filament changes with time, it always
remains visible from the inner to the outer penumbral boundary. We note that
the pixels with larger values of the linear polarization (marked with orange
contours) coincide with the positions of EC A. This behavior is also observed
in the other ECs. Hence, in addition to higher Doppler velocities, ECs display
larger linear polarization signals than the intra-spines along which they
move.

The polarization signals provide a rough estimate of the magnetic field
inclination under some restrictive assumptions: if the pixel is occupied by a
laterally homogeneous atmosphere, the lines are fully split, and the
orientation of the magnetic field remains constant along the LOS, a higher
inclination to the LOS implies larger $L$ and smaller $V$ signals, i.e.,
larger $L/V$ ratios. Thus, Fig.~\ref{fig:filaments} suggest that the intra-spines 
have more inclined magnetic fields than their surroundings, while 
Fig.~\ref{fig:ec_filament} could indicate that the ECs are the most inclined
structures inside them. However, the different behavior of the observables (in
particular Stokes $V$) emphasizes the need of inversions for an unambiguous
interpretation of the measurements.

\section{Periodicity}
\label{sec:periodicity}

As already mentioned, the Evershed flow displays variations on time scales of
8--25 min.  The quasi-periodic behavior is produced by ECs moving radially
outwards along intra-spines. The periodic times scales of 15~min are
associated with the appearance \citep{1994ApJ...430..413S} or
appearance/disappearance \citep{1994A&A...290..972R} of ECs in isolated
filaments. A precise estimation of the frequency of this repetitive behavior
requires a Fourier analysis of the data sets. Looking at
Figs.~\ref{fig:30general_1}--\ref{fig:30general_2} it is obvious that the
orientation of the filaments changes with time. Hence, one should carefully
select different paths for each map, otherwise the power spectrum will be
contaminated with information from spatial points outside the filament and
will show artificial peaks. Given its difficulties, we do not pursue this kind
of analysis here. However, a visual inspection of the maps provides valuable
information about the periodic appearance of ECs.

In Figs.~\ref{fig:30general_1} and \ref{fig:30general_2} one can observe a
repetitive behavior associated with the appearance of ECs along filament
$\#2$. Its time scale is about 15~min. A cycle begins with EC A at $t=3.9$
min. This EC arrives at the penumbral border around 15 minutes later. At that
moment, another EC appears in the mid-penumbra (EC B) and a new cycle
begins. After EC B reaches the outer penumbral edge, EC C appears and another
cycle starts.  New cycles take place in the filament with ECs D and F. This
behavior is irregular and does not happen during the whole time sequence.  For
filament $\#1$ the periodic time scale seems to be of the order of 40~min.
This is the time gap between the appearance of ECs E, G, and I (third column
of Table~\ref{tab:ecs_prop1}). Filament $\#3$ is visible during the second
half of the sequence. Only two ECs are observed to follow this filament, and
the time between their appearances is around 23 min. Therefore, our
observations suggest the existence of different time scales associated with
different filaments.  Longer time series mapping larger regions of the
penumbra are required to confirm or discard this possibility.

\section{Spectropolarimetric properties of ECs}
\label{sec:obs_prop}

In this Section we determine the typical Doppler velocities, $L/V$ ratios, and
Stokes $V$ area asymmetries associated with ECs. To simplify the analysis, a
single value of the parameters is calculated for each EC as follows.  We
compute the spatial average of the line parameters in each of the maps where a
given EC is observed, and then take the maximum of these average values.

Table~\ref{tab:ecs_prop2} shows the results for the spectral lines
with higher sensitivity to magnetic fields (\ion{Fe}{i} 1564.85~nm and
\ion{Fe}{i} 630.25 nm). The maximum unsigned blueshifts are about
3.2~\kms~and 2.2~\kms~for \ion{Fe}{i} 1564.85~nm and \ion{Fe}{i}
630.25~nm in the June 30 time sequence, and 2.1~\kms~and 1.3~\kms~in
the 1 July data set. These velocities are lower than the values
reported by \cite{1994ApJ...430..413S} and
\cite{1994A&A...290..972R}. The reason is that we calculate an average
over a group of pixels, while they used velocities from individual
pixels.

The mean linear-to-circular polarization ratios are of the order of 0.47 and
0.22 in \ion{Fe}{i} 1564.85~nm and \ion{Fe}{i} 630.25~nm for the June 30
observations, and around 0.81 and 0.37 for the July 1 data set. Since ECs 
show much larger $L/V$ ratios in the infrared, vector magnetographs working at
infrared wavelengths would detect them more easily than visible magnetographs.

The Stokes $V$ area asymmetry of \ion{Fe}{i} 1564.85~nm, $\delta A_{1564.85}$,
amounts to 5.9$\%$ on June 30 and 5.4$\%$ on July 1, while $\delta A_{630.25}$
is around 4.4$\%$ and 3.2$\%$ on June 30 and July 1, respectively. For some
ECs, the \ion{Fe}{i} 630.25~nm area asymmetry is of the order of its standard
deviation, suggesting that $\delta A_{630.25}$ has a strong dependence on
radial distance.

The large variations of the Doppler shifts and $L/V$ ratios of the ECs
as observed on two consecutive days are consistent with the idea that
the flow and magnetic field are almost horizontal to the solar
surface. The sunspot was nearer to disk center on July 1. If the
velocity and magnetic field vectors are horizontal to the solar
surface, their projections to the LOS would cause smaller blueshifts, 
larger linear polarization signals, and smaller circular polarization 
signals (i.e., larger $L/V$ ratios) as the spot approaches the disk center.
This explains why the velocities are higher on June 30 while the $L/V$
ratios are larger on July 1.

\section{Comparison of line parameters in ECs and intra-spines} 
\label{sec:comp_obs_prop}

\begin{figure}
\begin{center}
\scalebox{0.35}{\includegraphics{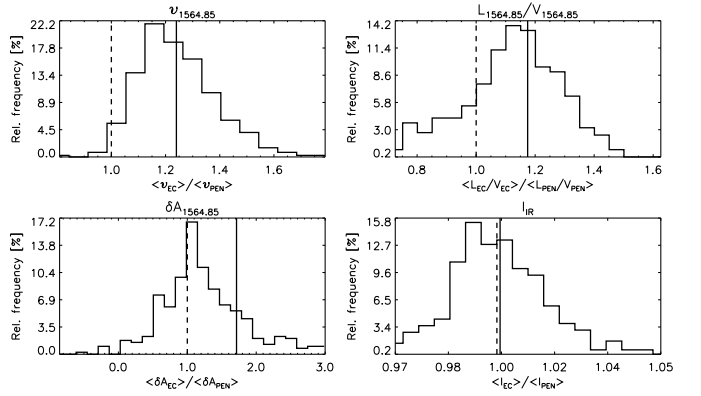}}
\caption{Normalized distribution of the ratio of Doppler velocities ($v$),
linear-to-circular polarization signals ($L/V$), Stokes $V$ area asymmetries
($\delta A$), and continuum intensities ($I$) of the ECs and the rest of
pixels of the intra-spines located at the same radial distance. The line
parameters are derived from \ion{Fe}{i} 1564.85~nm.  The vertical solid lines
represent the mean of the distributions.  The histograms correspond to the
June 30 data set.
\label{fig:30hist_comp_ir}}
\end{center}
\end{figure}

\begin{figure}
\begin{center}
\scalebox{0.35}{\includegraphics{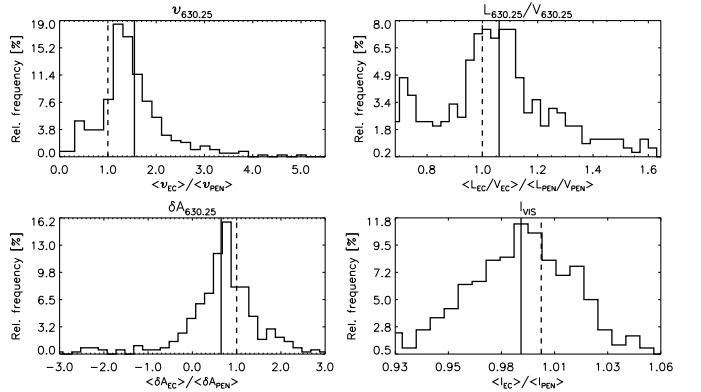}}
\caption{Same as Fig.~\ref{fig:30hist_comp_ir} for \ion{Fe}{i} 630.25
nm.\label{fig:30hist_comp_vis}}
\end{center}
\end{figure}

In a sunspot penumbra, the physical properties are known to vary radially from
the umbra to the outer penumbral boundary. For this reason, we compare the
observational properties of the ECs and those of the intra-spines as follows.
We calculate the mean line parameters at each radial distance for the pixels
that form the ECs and divide them by the mean parameters of the pixels inside 
the intra-spines which are at the same radial distance. This calculation is 
done for each map separately.  Figures \ref{fig:30hist_comp_ir} and 
\ref{fig:30hist_comp_vis} show histograms with the results of the analysis 
for the \ion{Fe}{i}~1564.8~nm and \ion{Fe}{i}~630.25~nm lines observed on June
30. The results for the July 1 data set are equivalent (not shown).

Interestingly, ECs exhibit larger Doppler velocities and $L/V$ ratios than 
the penumbral intra-spines in both \ion{Fe}{i} 630.25~nm and \ion{Fe}{i}
1564.85~nm. Hence, our measurements suggest that ECs possess the most inclined
magnetic fields of the penumbra. In Paper II we will confirm or disprove this
conjecture by inverting the observed spectra.

ECs also show larger \ion{Fe}{i} 1564.85~nm area asymmetries than the
intra-spines. This might be the result of larger gradients of the physical
quantities along the LOS and/or enhanced visibility of some magnetic
components of the penumbra in the ECs.  By contrast, $\delta A_{630.25}$ is
smaller than the intra-spine average inside the ECs. This opposite behavior is
surprising and could be due to the different heights of formation of the two
spectral lines.

\begin{figure}
\begin{center}
\scalebox{.59}{\includegraphics[bb=90 140 522 659,clip]{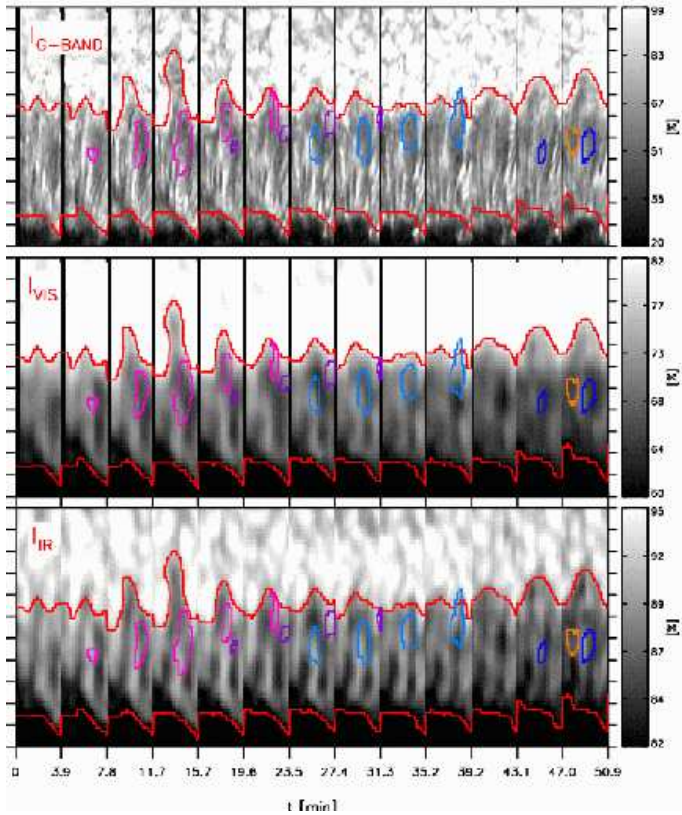}}
\caption{From {\em top} to {\em bottom}: Cospatial speckle-reconstructed 
G-band filtergrams, continuum intensity maps at 630~nm, and continuum
intensity maps at 1565~nm. Color contours outline the ECs.
\label{fig:conts}}
\end{center}
\end{figure}

The relation between ECs and continuum intensity is not well established in
the literature. \cite{1994ApJ...430..413S} and \cite{2003A&A...397..757R}
report that ECs are mainly correlated with locally bright continuum features,
while \cite{1994A&A...290..972R} and \cite{2003ApJ...584..509G} do not find
any correlation. Just by visual inspection of the observations it is difficult
to discriminate if ECs are bright or dark continuum features. In the intensity
histogram of Fig.~\ref{fig:30hist_comp_vis}, the ECs seem to show a preference
for locally dark continuum intensities at 630 nm. A similar trend is observed
in the continuum at 1565~nm (Fig.~\ref{fig:30hist_comp_ir}), but less
pronounced. These results have to be taken with caution because of the
existence of unresolved structures at the angular resolution of our
observations ($\sim \!0.6''-0.7''$). In Fig.~\ref{fig:conts} we display
co-spatial maps of speckle-reconstructed G-band intensity, continuum
intensity at 630~nm, and continuum intensity at 1565~nm. The figure
demonstrates that what we detect as dark/bright features in the
spectropolarimetric data are indeed a mixture of both in the G-band
filtergrams. As the ECs migrate outward, the dark/bright features inside 
them change, but in general the G-band intensities are locally darker at 
the position of the ECs. Spectropolarimetric observations at higher spatial
resolution are therefore required to draw definite conclusions on the
brightness of ECs.

Finally, we note that the histograms of Figs.~\ref{fig:30hist_comp_ir} and
\ref{fig:30hist_comp_vis} provide information about the average line
parameters of the ECs {\em independently of their position in the
penumbra}. These properties can change with radial distance. For example,
\cite{2005AN....326..301S} suggested that flow filaments link bright and
dark continuum features at different radial distances.  Since ECs move along
intra-spines (i.e., flow filaments), we may expect them to be associated with
locally bright features in the inner penumbra. This conjecture will be
examined in Sect.~\ref{sec:rad_obs_prop}.

\section{Dependence of Doppler velocities on height}
\label{sec:vlos_height}

\begin{figure}
\begin{center}
\scalebox{0.41}{\includegraphics{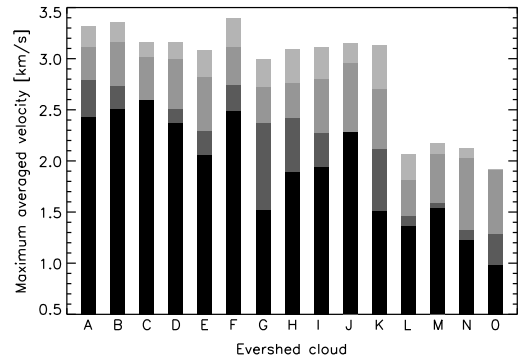}}
\caption{Maximum Doppler velocities of the observed ECs as derived from the
Stokes $V$ zero-crossing wavelengths. Maximum blueshifts for \ion{Fe}{i}
1564.85~nm, \ion{Fe}{i} 1565.28~nm, \ion{Ti}{i} 630.38~nm, and \ion{Fe}{i}
630.25 nm are represented with different shades of gray, from light to 
dark. Note that the blueshifts are unsigned.\label{fig:vlos_max}}
\end{center}
\end{figure}

The magnitude of the Doppler velocity depends on the spectral line chosen to
calculate it (cf.\ the first two columns of
Table~\ref{tab:ecs_prop2}). Figure~\ref{fig:vlos_max} shows the maximum
unsigned Doppler velocities of the ECs inferred from each spectral line.
There we see that the velocities indicated by the infrared lines are the
largest ones. The IR lines of the set are formed deeper than the others. This
implies that the Evershed flow is more conspicuous in deep atmospheric layers,
in agreement with previous findings \citep{1964ApNr....8..205M,
1997SoPh..171..331B, 2001ApJ...547.1148W, 2003A&A...410..695M, 2004A&A...415..731S,
2006A&A...453.1117B}.  The different velocities inferred
from the different spectral lines may be the result of (a) a change of the
modulus of the velocity vector with height; (b) a change of the inclination of
the velocity vector with height; or (c) a combination of both effects.

\section{Proper motions vs Doppler velocities}
\label{sec:motions_vel_correla}

It is of interest to compare the proper motions of the ECs along the LOS
(calculated as the motion along the LS multiplied by the sine of the
heliocentric angle) with their maximum Doppler velocities. If the blueshifts
associated with ECs in the center-side penumbra were produced by proper
motions, then both quantities should be similar in magnitude.

\begin{figure}
\begin{center}
\scalebox{0.35}{\includegraphics{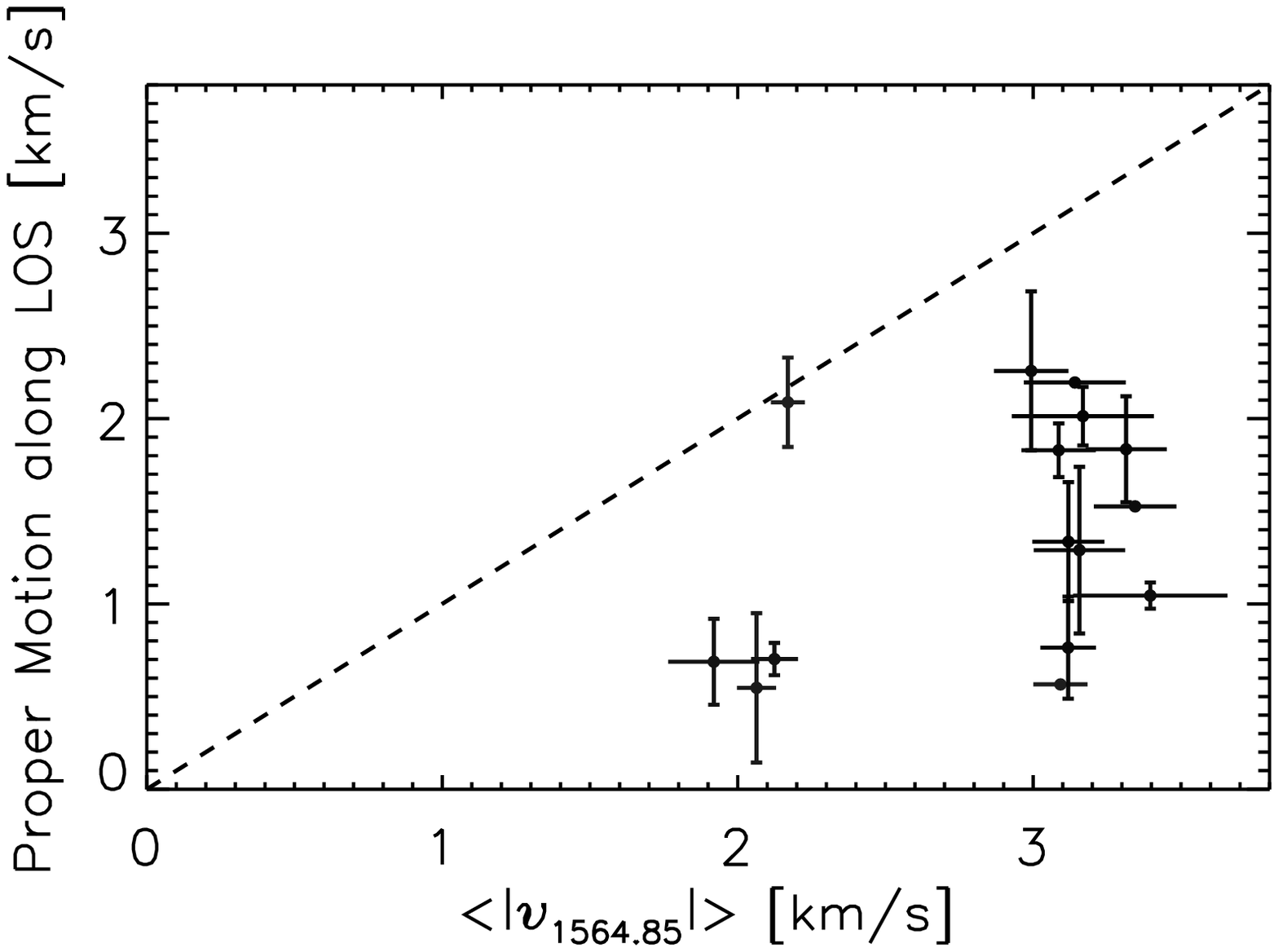}}
\scalebox{0.35}{\includegraphics{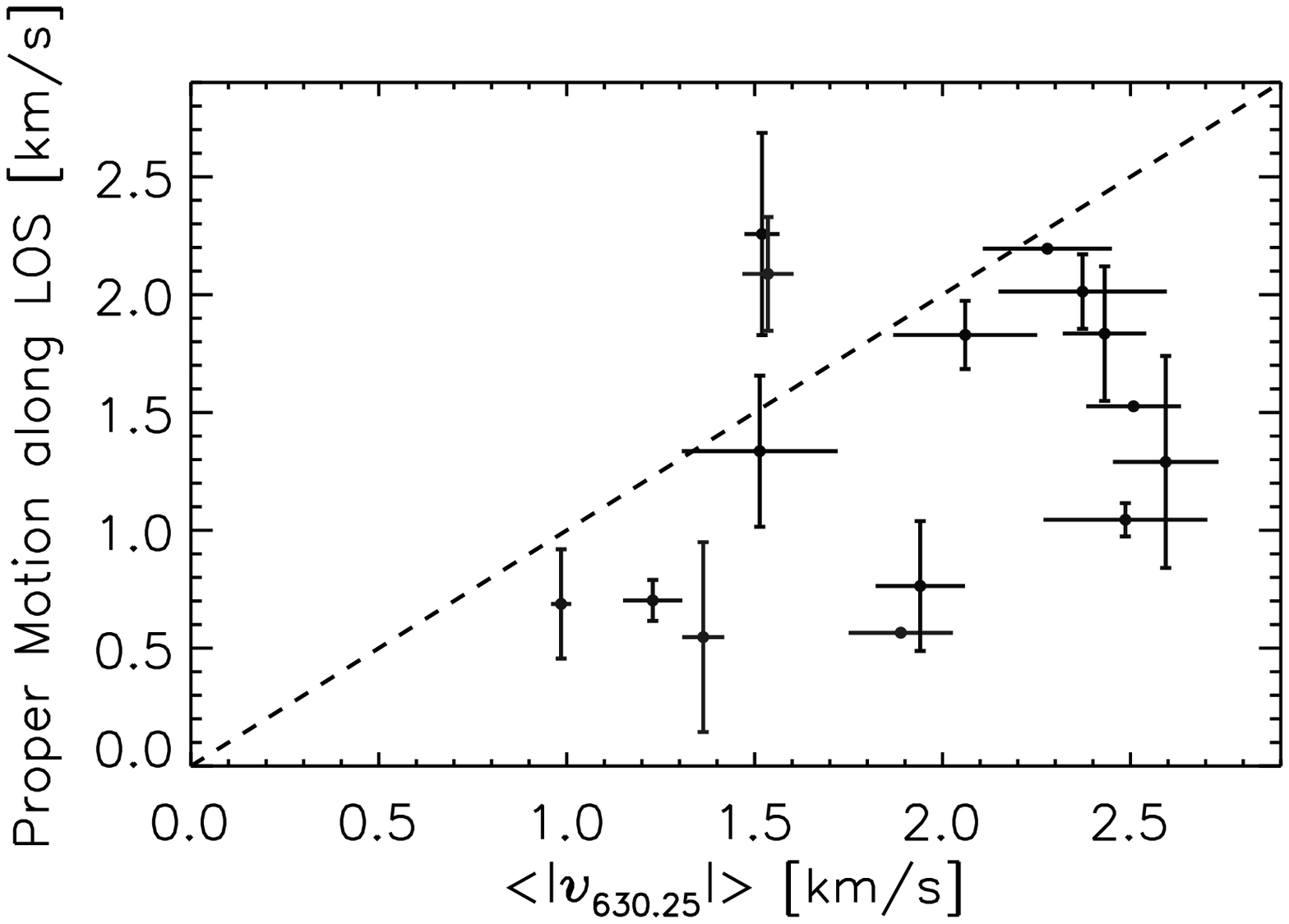}}
\caption{Proper motions along the LOS vs maximum Doppler velocities 
from \ion{Fe}{i} 1565.28~nm ({\em left}) and \ion{Fe}{i} 630.25~nm ({\em
right}) for the different ECs. Circles represent the maximum average
Doppler velocities. The horizontal bars represent their standard
deviations. The vertical bars are the errors in the determination of
proper motions.
\label{fig:vel_prop}}
\end{center}
\end{figure}

Figure~\ref{fig:vel_prop} plots the LOS projection of the propagation
velocities of the ECs vs the maximum Doppler velocities inferred from
\ion{Fe}{i} 1564.85~nm and \ion{Fe}{i} 630.25~nm. The horizontal bars indicate
the standard deviation of the Doppler velocities within each contour, and the
vertical bars the errors in the determination of propagation velocities. The
observed LOS velocities are systematically higher than the LOS component of
the proper motion. This result agrees with the findings of
\cite{1994ApJ...430..413S} and \cite{1994A&A...290..972R} and suggests that
the Evershed flow cannot only be the manifestation of the radial propagation
of ECs across the penumbra. The observed behavior is also compatible with the 
sea-serpent scenario described by \cite{2002AN....323..303S}, in which magnetic 
flux loops slowly migrate outward with high flow velocity inside them.

\section{Evolution of line parameters with radial distance}
\label{sec:rad_obs_prop}

\begin{figure}
\begin{center}
\scalebox{0.52}{\includegraphics{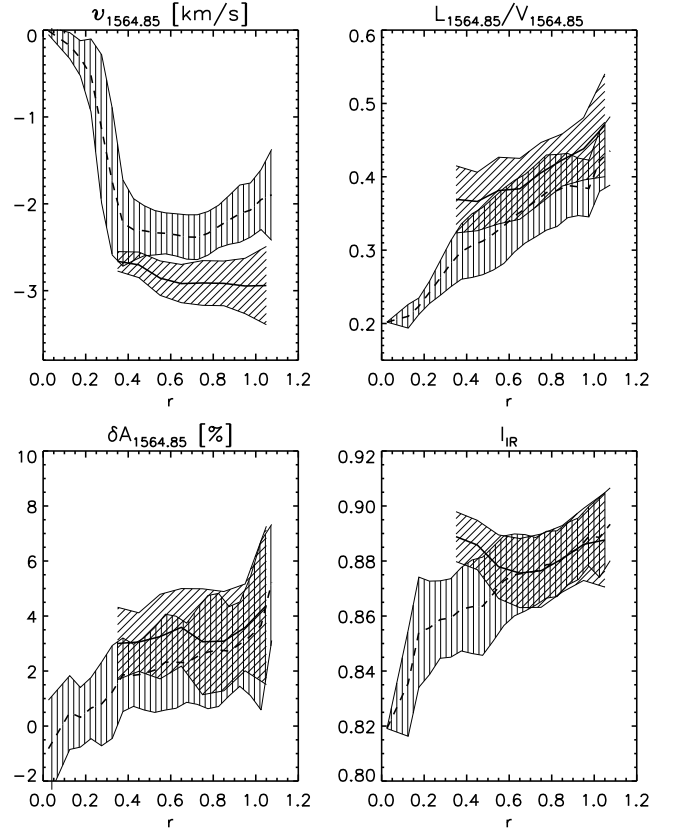}}
\caption{Average Stokes $V$ zero-crossing velocity, ratio of the total linear
and circular polarization, area asymmetry, and continuum intensity of
\ion{Fe}{i} 1564.85~nm ({\em top}) for pixels inside the ECs (solid line)
and the rest of pixels of the intra-spines (dashed line) as a function of radial
distance, for the time sequence observed on June 30.  Shaded areas indicate the
standard deviations around the mean value.
\label{fig:30evol_ecs_ir}} 
\end{center}
\end{figure}

\begin{figure}
\begin{center}
\scalebox{0.52}{\includegraphics{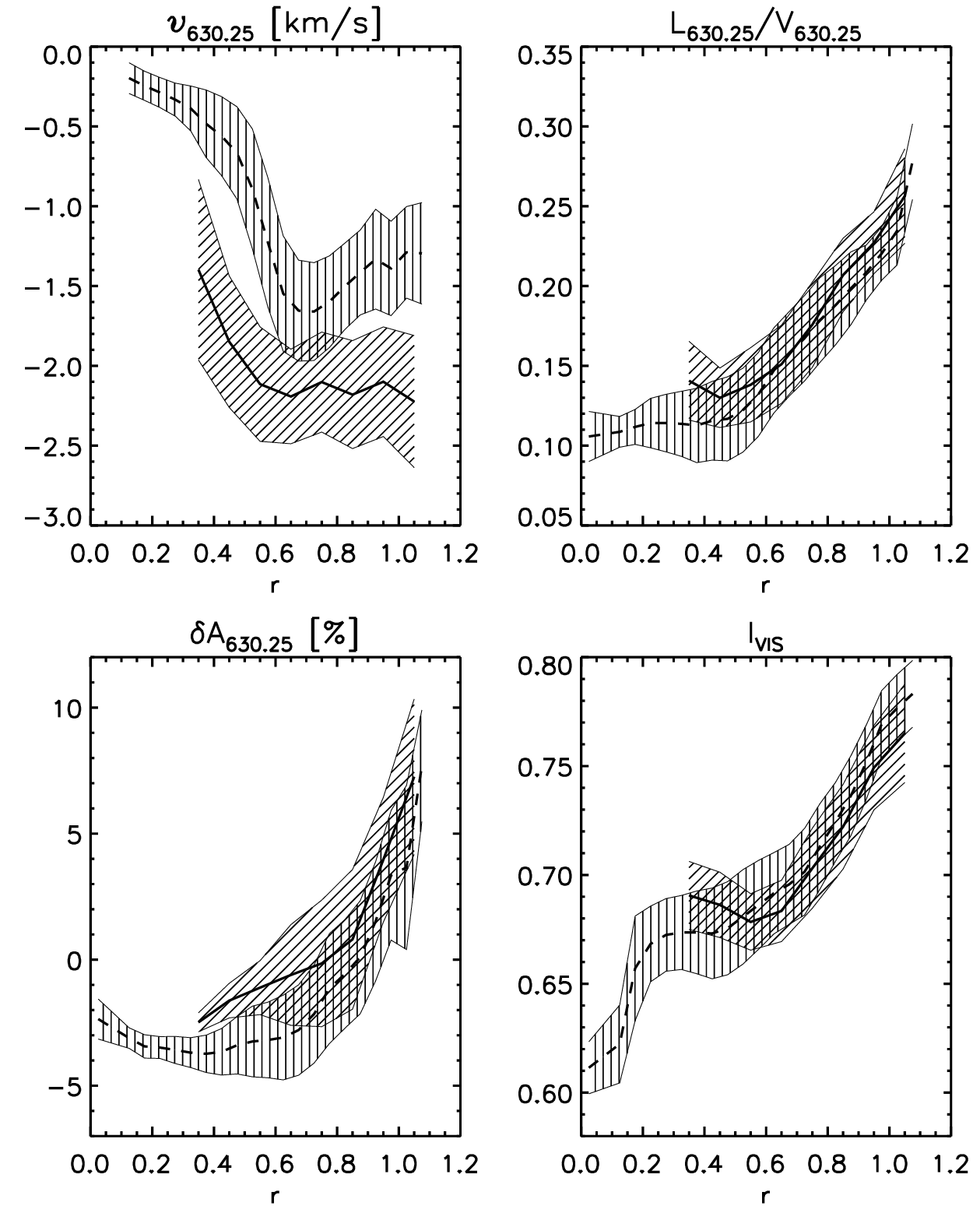}}
\caption{Same as Fig.~\ref{fig:30evol_ecs_ir} for the quantities
inferred from \ion{Fe}{i} 630.25 nm.
\label{fig:30evol_ecs_vis}} 
\end{center}
\end{figure}

In Sect.~\ref{sec:obs_prop} we compared the line parameters of ECs and
intra-spines disregarding the fact that there is a radial variation of the
physical conditions in the penumbra. Here we study how the line parameters
change with radial distance. To that aim, we average the parameters of pixels
inside the ECs which are at the same radial distance, and construct a curve of
their variation with $r$. We note that those curves represent the average
evolution of the ECs in the penumbra. To compare the behavior of the ECs and
intra-spines, we do the same calculation for pixels inside the intra-spines
excluding those of the ECs.

Figures~\ref{fig:30evol_ecs_ir} and \ref{fig:30evol_ecs_vis} show the results
of this analysis for \ion{Fe}{i} 1564.85~nm and \ion{Fe}{i} 630.25~nm,
respectively.  As can be seen, the Doppler velocities are greater inside the
ECs at all radial distances. Also interesting is the different shape of the
velocity curves. The velocity always increases within the ECs as they migrate
outward, from $v_{\rm 630.25}=-1.8$ \kms~and $v_{\rm 1564.85}=-2.8$ \kms~at
$r=0.5$ to $v_{\rm 630.25}=-2.2$ \kms~and $v_{\rm 1564.85}=-3.0$ \kms~at
$r=1$.  In the intra-spines, however, the Doppler velocity slightly decreases
from $r \sim 0.6$ outward.

The ratio of linear-to-circular polarization is larger inside the ECs at all
distances, which is more obvious for the IR line.  In addition, both the ECs
and the intra-spines display larger $L/V$ values at larger radial
distances. For the ECs, the $L/V$ ratio increases from $(L/V)_{630.25}=0.14$
at $r=0.5$ and $(L/V)_{1564.85}=0.38$ up to $(L/V)_{630.25}=0.24$ and
$(L/V)_{1564.85}=0.47$ at $r=1$.

The area asymmetry $\delta A$ of the two lines also increases monotonically
with radial distance, showing larger values within the ECs than in the
intra-spines. The area asymmetries change from $\delta A_{630.25}=-1.5\%$ and
$\delta A_{1564.85}$ $=4.5\%$ at $r=0.5$ to $\delta A_{630.25}=6.2\%$ and
$\delta A_{1564.85}=7.2\%$ at $r=1$.

The continuum intensity curves display an interesting variation with radial
distance. Both $I_{\rm IR}$ and $I_{\rm VIS}$ are greater than the intra-spine
average inside the ECs at the radial distances where they appear ($r\sim0.4$).
From the mid to the outer penumbra, however, the continuum intensities of ECs and
intra-spines do not exhibit significantly different values.

\section{Disappearance of type I ECs}
\label{sec:ec_dissapear}

Most of the type {\rm I} ECs observed on June 30 disappear suddenly in the
penumbra, near the outer sunspot boundary.  This indicates that the time they
take to vanish is usually shorter than the cadence of the observations (3.9
min on June 30). There is, however, an interesting case in which a type {\rm
I} EC is seen to disappear.

\begin{figure}
\begin{center}
\scalebox{0.6}{\includegraphics{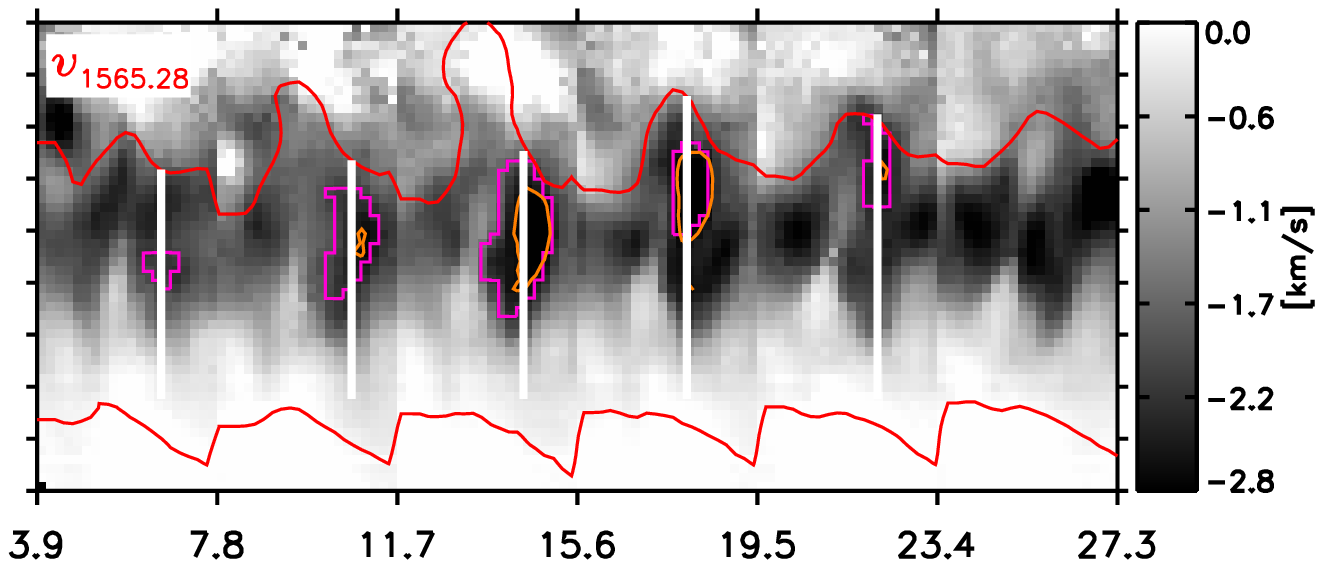}}
\caption{Maps of Stokes $V$ zero-crossing velocity of \ion{Fe}{i} 1565.28~nm
as a function of time. Pink contours mark EC A. Orange contours enclose pixels
with $v_{630.25}$ and $v_{1565.28}$ lower than $-2.3$~\kms~and $-2.7$~\kms~,
respectively. The vertical lines mark a radial cut passing through the
EC. Each tickmark in the $y$-axis represents 1\arcsec. \label{fig:ec_a_decay}}
\end{center}
\end{figure}

\begin{figure*}
\begin{center}
\scalebox{0.507}{\includegraphics{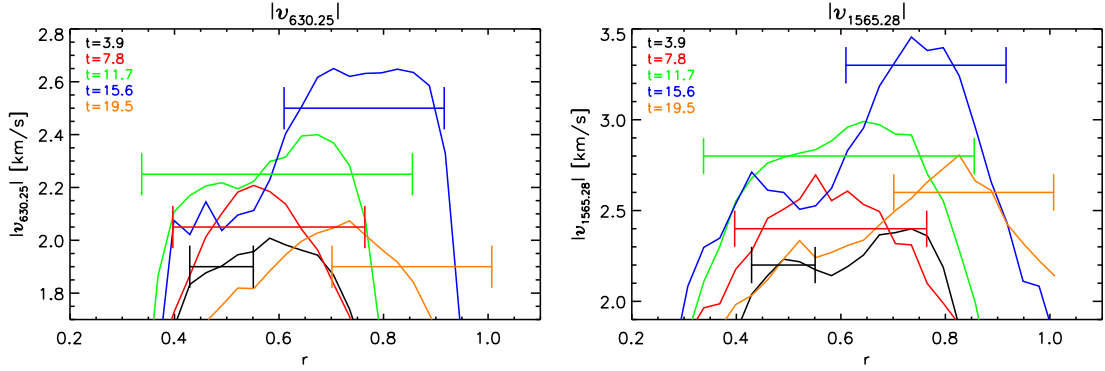}}
\caption{Variation of $|v_{630.25}|$ and $|v_{1565.28}|$ along the
radial cuts shown in Fig.~\ref{fig:ec_a_decay}. The colors indicate
different times and the horizontal lines mark the upper and lower
edges of EC A.\label{fig:var_ec_decay}}
\end{center}
\end{figure*}

\begin{figure*}
\begin{center}
\scalebox{0.84}{\includegraphics{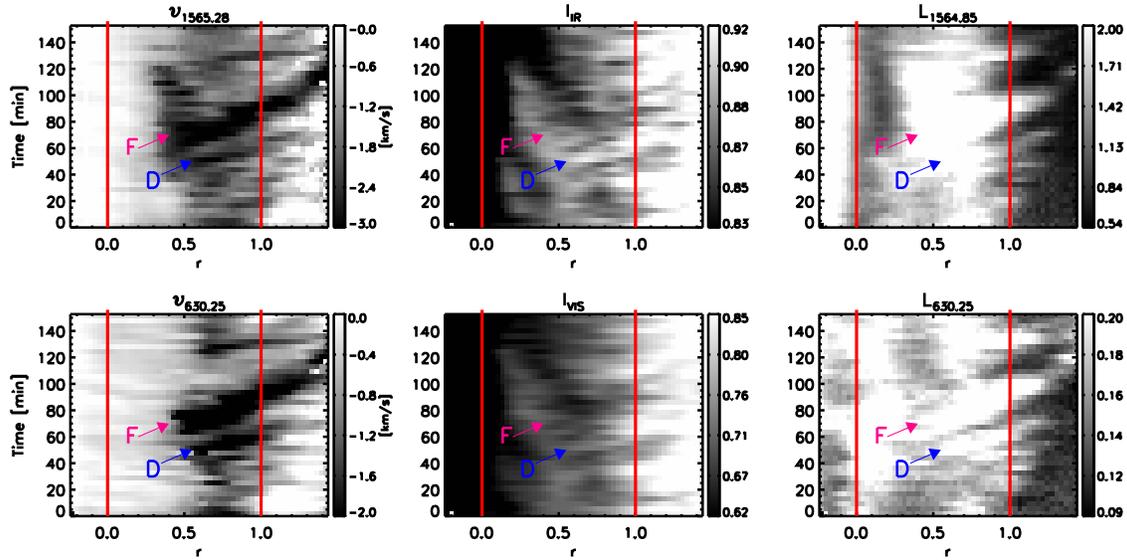}}
\caption{Temporal evolution of line parameters along intra-spine \#2 (cf.\ the
green radial cut in the first panel of Fig.~\ref{fig:filaments}).  {\em Top:}
Stokes $V$ zero-crossing velocity of \ion{Fe}{i}~1565.28~nm (left), continuum
intensity at 1565~nm (center), and total linear polarization of \ion{Fe}{i}
1564.85~nm (right). {\em Bottom:} Stokes $V$ zero-crossing velocity of
\ion{Fe}{i} 630.25~nm (left), continuum intensity at 630.2~nm (center), and
total linear polarization of \ion{Fe}{i} 630.25 nm (right). The $x$-axis 
gives the average penumbral radial distance and the $y$-axis displays 
time. The arrows indicate the positions of ECs D and F.
\label{fig:30_front}}
\end{center}
\end{figure*}

Figure~\ref{fig:ec_a_decay} shows the evolution of the Doppler velocities of
EC A derived from \ion{Fe}{i} 1565.28~nm. As can be seen, EC A has vanished at
$t=23.4$~min (there is no dark patch where one would expect to find it, just
outside of the penumbra). In order to quantify the evolution of the velocities
in the EC, we choose radial cuts going from the inner to the outer penumbral
boundary passing through the middle of the EC in each map. These cuts are
indicated by the vertical solid lines in Fig.~\ref{fig:ec_a_decay}. The
variation of the Doppler velocities along the cuts is shown in
Fig.~\ref{fig:var_ec_decay}.  Colors indicate different times and the
horizontal lines mark the upper and lower edges of the EC. In agreement with
the behavior indicated by Figs.~\ref{fig:30evol_ecs_ir} and
\ref{fig:30evol_ecs_vis}, the EC velocities increase with radial distance
until $t=15.6$~min.  Then, there is a strong drop in the velocities at
$t=19.5$~min, after which the EC is no longer seen. This behavior is also
observed for EC C and must be general for type {\rm I} ECs.

\section{Evolution of type II ECs beyond the outer penumbral boundary}
\label{sec:out_pen_gen}

\cite{1994ApJ...430..413S} and \cite{1994A&A...290..972R} reported that some
ECs penetrate slightly beyond the outer penumbral boundary before dissolving.
On June 30 and July 1, four ECs crossed the edge of AR 10781. Here we describe
the evolution of these type II ECs in the sunspot moat.

\subsection{Modification of the outer edge of the penumbra by ECs}

In Figure~\ref{fig:filaments} we presented line parameter maps for
filament~\#2. ECs D and F moved along this
intra-spine. Figure~\ref{fig:30_front} displays the temporal evolution of the
Doppler velocity, the continuum intensity, and the total linear polarization
as a function of the mean radial distance for a cut along the intra-spine. The
three signals are strongly correlated near the outer penumbral boundary, which
becomes more evident when an EC reaches the quiet photosphere. This happens at
$t=58.7$~min for EC D and $t=97.7$~min for EC F.  When an EC penetrates into
the quiet photosphere, the outer penumbra does the same: the velocity and
linear polarization increase and the continuum intensity decreases in the
quiet photosphere, i.e., the penumbra grows at that position. Such a behavior
has already been reported by \cite{1994ApJ...430..413S} for the intensity and
velocity signals.

\subsection{Disappearance of type II ECs in the moat}
\label{sec:ecs_dis}

The evolution of ECs D and F outside of the spot can be followed in Figs.~5
and 6. EC D (pink contours) crossed the penumbral boundary between $t=93.8$
and $t=97.7$ min and its proper motion outside the spot was $\sim$~1.9~km~s$^{-1}$.  
Anomalous Stokes $V$ profiles appear where one would locate the EC at 
$t=109.6$ min. We cannot calculate Doppler velocities from the Stokes $V$ 
profiles observed in those pixels.  However, the EC is easily detected in 
the velocity maps computed from the Doppler shifts of Stokes $I$ (not shown). 
The evolution of EC F can be tracked until $t=109.6$ min. Thus, EC F 
disappeared as a velocity structure $11.9$~min after crossing the spot 
boundary.

The evolution of ECs M and N outside the spot is shown in
Figs.~\ref{fig:01general_1} and \ref{fig:01general_2}.  EC M is only
visible in two maps before it goes out of the FOV. EC N disappeared 2$\farcs$1
from the spot boundary.

The phenomenological properties of ECs D, F, M, and N in the moat are
summarized in Table~\ref{tab:ecs_outside}. The behavior of type II ECs outside
the spot is very similar: they quickly dissolve ($\sim$13.8~min) without
reaching large radial distances from the outer penumbral boundary
($\sim$1\farcs9).  Inside the spot ECs show up as elongated structures, but
type II ECs adopt a roundish shape when they enter the moat. Moreover, their
proper motions are reduced relative to those they exhibit in the penumbra
(compare Tables~\ref{tab:ecs_outside} and 1).

As discussed by \cite{2006ApJ...649L..41C}, ECs D and F became moving magnetic
features (MMFs) in the sunspot moat. This suggests that type II ECs are the
precursors of at least some MMFs. The relation between MMFs and the magnetic
field of the penumbra has been studied by \cite{2005ApJ...632.1176S}, 
\cite{2007kubo}, and \cite{cabrera.2007b}.

\section{Summary and conclusions}
\label{sec:conclusions_cap5}

We have presented a detailed description of the temporal evolution of the
Evershed flow including, for the first time, its polarization signatures. 
Our main findings can be summarized as follows:

\begin{enumerate}

\item We identify fifteen Evershed clouds (ECs) that appear in the mid
penumbra. They propagate with speeds of $\sim$2.6~\kms~along penumbral
filaments characterized by larger values of the linear polarization,
linear-to-circular polarization ratios, and Doppler velocities than their
surroundings. These filaments can be associated with the penumbral
intra-spines described by \cite{1993ApJ...418..928L}.

\begin{table}
\begin{center}
\tabcolsep .7em
\caption{Phenomenological properties of type II ECs in the moat. The time and
distance from the outer penumbral boundary where the ECs vanish as velocity
structures are given in the second and third columns. The fourth column
displays the time each of them survived in the quiet photosphere. The 
proper motions of the ECs in the moat are given in the fifth column. The 
last two columns display their maximum lengths and widths, corrected for LOS
effects, outside the penumbra. The parentheses indicate that EC M leaves the
FOV before disappearing.\label{tab:ecs_outside}}
\begin{tabular}{l r c c c r r}
\hline
\hline
\multicolumn{1}{l}{EC} &
\multicolumn{1}{c}{$t_{\rm dis}$} &
\multicolumn{1}{c}{$d_{\rm dis}$} &
\multicolumn{1}{c}{$\tau_{\rm ph}$} &
\multicolumn{1}{c}{$v_{\rm prop}$} & \multicolumn{1}{c}{Length} &
\multicolumn{1}{c}{Width}\\& [min] & [\arcsec] &[min] & [\kms] & [km] & [km] \\
\hline
D & 74.3  & 1.8 & 15.6 & $0.6 \pm 0.3$   & 1300 & 700  \\ 
F & 109.6 & 1.8 & 11.9 & $1.9 \pm 0.4$   & 1000 & 1200 \\ 
(M) &     &     &      & $3.5 \pm 0.0$ & 900  & 400  \\ 
N & 57.8  & 2.1 & 14.0 & $1.3 \pm 0.1$   & 900  & 600  \\ 
\hline
{\bf Mean} & & {\bf 1.9} & {\bf 13.8} &{\bf 1.8} & {\bf 1000} & {\bf 700} \\
\hline
\end{tabular}
\end{center}
\end{table}

\item ECs can be classified in two types: (a) ECs that disappear in the outer
penumbra (type {\rm I}); and (b) ECs that cross the visible border of the spot
and enter the sunspot moat (type {\rm II}). Most of the observed ECs belong to
type {\rm I}.

\item The motions of ECs in the penumbra are mainly radial.

\item ECs have greater lengths ($\sim$1700~km) than widths ($\sim$700~km), 
and their sizes increase as they migrate outward.

\item Our observations suggest the existence of different time scales
associated with the appearance of the ECs. We find periodic behaviors of 
the order of 15 min and 40 min in two different intra-spines.

\item ECs display larger Doppler velocities and $L/V$ ratios in the 
infrared lines of the set. This implies that: (a) ECs are more conspicuous 
in deeper atmospheric layers; and (b) the use of infrared lines would be 
advantageous to detect them in vector magnetograms.

\item ECs possess greater Doppler velocities, $L/V$ ratios, and area
asymmetries than the intra-spines along which they migrate outward. This might
imply that the magnetic and dynamic configuration of the intra-spines are
modified by the passage of ECs. Assuming that the velocity and magnetic field
vectors are aligned \citep{2003A&A...403L..47B}, the enhancement of Doppler
velocity displayed by the ECs can only be produced by an increase in 
the modulus of the velocity vector.

\item The Doppler velocities observed in ECs are larger when the sunspot 
is farther from disk center, while their $L/V$ ratios are smaller. These
results are consistent with the idea that the flow and the magnetic
field are almost horizontal to the solar surface.

\item ECs are brighter than intra-spines in the inner penumbra. The brightness
excess is no longer seen in the mid and outer penumbra.

\item The Doppler velocities of the ECs are systematically greater than 
the LOS projection of their proper motions, assumed horizontal to the 
solar surface.

\item The blueshifts, linear-to-circular polarization ratios, and area
asymmetries of the ECs increase with radial distance. In the intra-spines,
$L/V$ and $\delta A$ display the same behavior, while the velocity decreases
from the middle to the outer penumbra.

\item The disappearance time scales of type {\rm I} ECs are shorter 
than the cadence of $\sim 4$~min of the observations.

\item The outer penumbral boundary is modified by the passage of the ECs that
reach the quiet photosphere. When a type II EC penetrates into the moat, the
outer penumbral boundary does the same and the intensity decreases in the
quiet photosphere.

\item Four ECs are seen to cross the visible border of the spot. Once they
enter the moat: (a) their propagation velocities are smaller than inside
the penumbra (0.9 vs 2.6~\kms); (b) they adopt a roundish shape; and (c) 
their blueshifts quickly decrease until they vanish. Type II ECs can be
detected as velocity structures in the quiet photosphere for only about 
14 min after crossing the outer penumbral boundary. They disappear relatively
close to the spot ($\sim$2\arcsec).

\item Some type II ECs become moving magnetic features in the sunspot moat 
\citep{2006ApJ...649L..41C}, but they do not show significant Doppler 
shifts.

\end{enumerate}

The line parameters derived from the polarization profiles give us valuable
information about the phenomenological properties of the ECs. However, the
interpretation of these observational parameters in terms of physical
quantities is not straightforward. For example, a larger value of the 
Doppler shift can be associated with a stronger Evershed flow, but 
also with a magnetic field which is less inclined to the LOS.

To remove these ambiguities, we have perform detailed Stokes inversions 
of the observed spectra. The results of the inversions, presented in 
Paper II of this series, will allow us to investigate the nature of the
EC phenomenon.

\begin{acknowledgements}
This work has been supported by the Spanish MEC under project
ESP2006-13030-C06-02 and {\em Programa Ram\'on y Cajal}. The German VTT is
operated by the Kiepenheuer-Institut f\"ur Sonnenphysik at the Observatorio
del Teide of the Instituto de Astrof\'{\i}sica de Canarias. The DOT is
operated by Utrecht University at the Observatorio del Roque de Los Muchachos
on La Palma, also of the Instituto de Astrof\'{\i}sica de Canarias.
\end{acknowledgements}

\bibliographystyle{aa}

\begin{thebibliography}
\expandafter\ifx\csname natexlab\endcsname\relax\def\natexlab#1{#1}\fi

\bibitem[{{Allende Prieto} {et~al.}(2004){Allende Prieto}, {Asplund}, \&
  {Fabiani Bendicho}}]{2004A&A...423.1109A}
{Allende Prieto}, C., {Asplund}, M., \& {Fabiani Bendicho}, P. 2004, \aap, 423,
  1109

\bibitem[{{Auer} \& {Heasley}(1978)}]{1978A&A....64...67A}
{Auer}, L.~H. \& {Heasley}, J.~N. 1978, \aap, 64, 67

\bibitem[{{Balthasar} {et~al.}(1997){Balthasar}, {Schmidt}, \&
  {Wiehr}}]{1997SoPh..171..331B}
{Balthasar}, H., {Schmidt}, W., \& {Wiehr}, E. 1997, \solphys, 171, 331

\bibitem[{{Beck} {et~al.}(2007){Beck}, {Bellot Rubio}, {Schlichenmaier}, \&
  {S{\"u}tterlin}}]{2007.beck_a}
{Beck}, C., {Bellot Rubio}, L., {Schlichenmaier}, R., \& {S{\"u}tterlin}, P.
  2007, \aap, in press, arXiv:0707.1232

\bibitem[{{Beck} {et~al.}(2005{\natexlab{a}}){Beck}, {Schlichenmaier},
  {Collados}, {Bellot Rubio}, \& {Kentischer}}]{2005A&A...443.1047B}
{Beck}, C., {Schlichenmaier}, R., {Collados}, M., {Bellot Rubio}, L., \&
  {Kentischer}, T. 2005{\natexlab{a}}, \aap, 443, 1047

\bibitem[{{Beck} {et~al.}(2005{\natexlab{b}}){Beck}, {Schmidt}, {Kentischer},
  \& {Elmore}}]{2005A&A...437.1159B}
{Beck}, C., {Schmidt}, W., {Kentischer}, T., \& {Elmore}, D.
  2005{\natexlab{b}}, \aap, 437, 1159

\bibitem[{{Beckers}(1977)}]{1977ApJ...213..900B}
{Beckers}, J.~M. 1977, \apj, 213, 900

\bibitem[{{Bellot Rubio} {et~al.}(2003){Bellot Rubio}, {Balthasar}, {Collados},
  \& {Schlichenmaier}}]{2003A&A...403L..47B}
{Bellot Rubio}, L.~R., {Balthasar}, H., {Collados}, M., \& {Schlichenmaier}, R.
  2003, \aap, 403, L47

\bibitem[{{Bellot Rubio} {et~al.}(2006){Bellot Rubio}, {Schlichenmaier}, \&
  {Tritschler}}]{2006A&A...453.1117B}
{Bellot Rubio}, L.~R., {Schlichenmaier}, R., \& {Tritschler}, A. 2006, \aap,
  453, 1117

\bibitem[{{Borrero} {et~al.}(2003){Borrero}, {Bellot Rubio}, {Barklem}, \& {del
  Toro Iniesta}}]{2003A&A...404..749B}
{Borrero}, J.~M., {Bellot Rubio}, L.~R., {Barklem}, P.~S., \& {del Toro
  Iniesta}, J.~C. 2003, \aap, 404, 749

\bibitem[{{Cabrera Solana}(2007)}]{cabrera.thesis}
{Cabrera Solana}, D. 2007, Ph.D.\ Thesis, University of Granada, Spain

\bibitem[{{Cabrera Solana} {et~al.}(2006){Cabrera Solana}, {Bellot Rubio},
  {Beck}, \& {del Toro Iniesta}}]{2006ApJ...649L..41C}
{Cabrera Solana}, D., {Bellot Rubio}, L.~R., {Beck}, C., \& {del Toro Iniesta},
  J.~C. 2006, \apjl, 649, L41

\bibitem[{{Cabrera Solana} {et~al.}(2008){Cabrera Solana}, {Bellot Rubio},
  {Beck}, \& {del Toro Iniesta}}]{cabrera.2007b}
{Cabrera Solana}, D., {Bellot Rubio}, L.~R., {Beck}, C., \& {del Toro Iniesta},
  J.~C. 2008, \aap, in preparation

\bibitem[{{Cabrera Solana} {et~al.}(2007){Cabrera Solana}, {Bellot Rubio},
  {Borrero}, \& {del Toro Iniesta}}]{cabrera.2007a}
{Cabrera Solana}, D., {Bellot Rubio}, L.~R., {Borrero}, J.~M., \& {del Toro
  Iniesta}, J.~C. 2007, \aap, submitted (Paper II)

\bibitem[{{Cabrera Solana} {et~al.}(2005){Cabrera Solana}, {Bellot Rubio}, \&
  {del Toro Iniesta}}]{2005A&A...439..687C}
{Cabrera Solana}, D., {Bellot Rubio}, L.~R., \& {del Toro Iniesta}, J.~C. 2005,
  \aap, 439, 687

\bibitem[{{Collados} {et~al.}(1999){Collados}, {Rodr{\'{\i}}guez Hidalgo},
  {Bellot Rubio}, {Ruiz Cobo}, \& {Soltau}}]{1999AGM....15..A13C}
{Collados}, M., {Rodr{\'{\i}}guez Hidalgo}, I., {Bellot Rubio}, L., {Ruiz
  Cobo}, B., \& {Soltau}, D. 1999, in Astronomische Gesellschaft Meeting
  Abstracts, ed. R.~E. {Schielicke}, 13

\bibitem[{{Collados}(2003)}]{2003SPIE.4843...55C}
{Collados}, M.~V. 2003, in SPIE 4843: Polarimetry in Astronomy, ed.
  S.~{Fineschi}, 55

\bibitem[{{del Toro Iniesta} {et~al.}(1994){del Toro Iniesta}, {Tarbell}, \&
  {Ruiz Cobo}}]{1994ApJ...436..400D}
{del Toro Iniesta}, J.~C., {Tarbell}, T.~D., \& {Ruiz Cobo}, B. 1994, \apj,
  436, 400

\bibitem[{{Georgakilas} \& {Christopoulou}(2003)}]{2003ApJ...584..509G}
{Georgakilas}, A.~A. \& {Christopoulou}, E.~B. 2003, \apj, 584, 509

\bibitem[{{Kubo} {et~al.}(2007){Kubo}, {Shimizu}, \& {Tsuneta}}]{2007kubo}
{Kubo}, M., {Shimizu}, T., \& {Tsuneta}, S. 2007, \apj, 659, 812

\bibitem[{{Landi degl'Innocenti}(1992)}]{1992soti.book...71D}
{Landi degl'Innocenti}, E. 1992, {Magnetic field measurements} (Solar
  Observations: Techniques and Interpretation), 71

\bibitem[{{Lites} {et~al.}(1993){Lites}, {Elmore}, {Seagraves}, \&
  {Skumanich}}]{1993ApJ...418..928L}
{Lites}, B.~W., {Elmore}, D.~F., {Seagraves}, P., \& {Skumanich}, A.~P. 1993,
  \apj, 418, 928

\bibitem[{{Maltby}(1964)}]{1964ApNr....8..205M}
{Maltby}, P. 1964, Astrophysica Norvegica, 8, 205

\bibitem[{{Mart{\'{\i}}nez Pillet} {et~al.}(1999){Mart{\'{\i}}nez Pillet},
  {Collados}, {S{\'a}nchez Almeida}, {Gonz{\'a}lez}, {Cruz-Lopez}, {Manescau},
  {Joven}, {Paez}, {Diaz}, {Feeney}, {S{\'a}nchez}, {Scharmer}, \&
  {Soltau}}]{1999ASPC..183..264M}
{Mart{\'{\i}}nez Pillet}, V., {Collados}, M., {S{\'a}nchez Almeida}, J.,
  {et~al.} 1999, in ASP Conf. Ser. 183: High Resolution Solar Physics: Theory,
  Observations, and Techniques, ed. T.~R. {Rimmele}, K.~S. {Balasubramaniam},
  \& R.~R. {Radick}, 264

\bibitem[{{Mathew} {et~al.}(2003){Mathew}, {Lagg}, {Solanki}, {Collados},
  {Borrero}, {Berdyugina}, {Krupp}, {Woch}, \&
  {Frutiger}}]{2003A&A...410..695M}
{Mathew}, S.~K., {Lagg}, A., {Solanki}, S.~K., {et~al.} 2003, \aap, 410, 695

\bibitem[{{Nave} {et~al.}(1994){Nave}, {Johansson}, {Learner}, {Thorne}, \&
  {Brault}}]{1994ApJS...94..221N}
{Nave}, G., {Johansson}, S., {Learner}, R.~C.~M., {Thorne}, A.~P., \& {Brault},
  J.~W. 1994, \apjs, 94, 221

\bibitem[{{Pierce} \& {Breckenridge}(1974)}]{1974kptp.book.....P}
{Pierce}, A.~K. \& {Breckenridge}, J.~B. 1974, {The Kitt Peak table of
  photographic solar spectrum wavelengths} (Kitt Peak National Observatory
  Contribution, Tucson: Kitt Peak National Observatory, 1973-1974)

\bibitem[{{Reardon}(2006)}]{2006SoPh..239..503R}
{Reardon}, K.~P. 2006, \solphys, 239, 503

\bibitem[{{Rezaei} {et~al.}(2006){Rezaei}, {Schlichenmaier}, {Beck}, \& {Bellot
  Rubio}}]{2006A&A...454..975R}
{Rezaei}, R., {Schlichenmaier}, R., {Beck}, C., \& {Bellot Rubio}, L.~R. 2006,
  \aap, 454, 975

\bibitem[{{Rimmele}(1994)}]{1994A&A...290..972R}
{Rimmele}, T.~R. 1994, \aap, 290, 972

\bibitem[{{Rouppe van der Voort}(2003)}]{2003A&A...397..757R}
{Rouppe van der Voort}, L.~H.~M. 2003, \aap, 397, 757

\bibitem[{{Sainz Dalda} \& {Mart{\'{\i}}nez
  Pillet}(2005)}]{2005ApJ...632.1176S}
{Sainz Dalda}, A. \& {Mart{\'{\i}}nez Pillet}, V. 2005, \apj, 632, 1176

\bibitem[{{S{\'a}nchez Almeida} \& {Lites}(1992)}]{1992ApJ...398..359S}
{S{\'a}nchez Almeida}, J. \& {Lites}, B.~W. 1992, \apj, 398, 359

\bibitem[{{Schlichenmaier}(2002)}]{2002AN....323..303S}
{Schlichenmaier}, R. 2002, Astronomische Nachrichten, 323, 303

\bibitem[{{Schlichenmaier} {et~al.}(2004){Schlichenmaier}, {Bellot Rubio}, \&
  {Tritschler}}]{2004A&A...415..731S}
{Schlichenmaier}, R., {Bellot Rubio}, L.~R., \& {Tritschler}, A. 2004, \aap,
  415, 731

\bibitem[{{Schlichenmaier} {et~al.}(2005){Schlichenmaier}, {Bellot Rubio}, \&
  {Tritschler}}]{2005AN....326..301S}
{Schlichenmaier}, R., {Bellot Rubio}, L.~R., \& {Tritschler}, A. 2005,
  Astronomische Nachrichten, 326, 301

\bibitem[{{Schlichenmaier} \& {Collados}(2002)}]{2002A&A...381..668S}
{Schlichenmaier}, R. \& {Collados}, M. 2002, \aap, 381, 668

\bibitem[{{Schmidt} {et~al.}(2003){Schmidt}, {Beck}, {Kentischer}, {Elmore}, \&
  {Lites}}]{2003AN....324..300S}
{Schmidt}, W., {Beck}, C., {Kentischer}, T., {Elmore}, D., \& {Lites}, B. 2003,
  Astronomische Nachrichten, 324, 300

\bibitem[{{Shine} {et~al.}(1994){Shine}, {Title}, {Tarbell}, {Smith}, {Frank},
  \& {Scharmer}}]{1994ApJ...430..413S}
{Shine}, R.~A., {Title}, A.~M., {Tarbell}, T.~D., {et~al.} 1994, \apj, 430, 413

\bibitem[{{Socas-Navarro} {et~al.}(2004){Socas-Navarro}, {Pillet}, {Sobotka},
  \& {V{\'a}zquez}}]{2004ApJ...614..448S}
{Socas-Navarro}, H., {Pillet}, V.~M., {Sobotka}, M., \& {V{\'a}zquez}, M. 2004,
  \apj, 614, 448

\bibitem[{{Solanki}(1993)}]{1993SSRv...63....1S}
{Solanki}, S.~K. 1993, Space Science Reviews, 63, 1

\bibitem[{{Solanki}(2003)}]{2003A&ARv..11..153S}
{Solanki}, S.~K. 2003, \aapr, 11, 153

\bibitem[{{Soltau} {et~al.}(2002){Soltau}, {Berkefeld}, {von der L{\"u}he},
  {W{\"o}ger}, \& {Schelenz}}]{2002AN....323..236S}
{Soltau}, D., {Berkefeld}, T., {von der L{\"u}he}, O., {W{\"o}ger}, F., \&
  {Schelenz}, T. 2002, Astronomische Nachrichten, 323, 236

\bibitem[{{Westendorp Plaza} {et~al.}(2001){Westendorp Plaza}, {del Toro
  Iniesta}, {Ruiz Cobo}, \& {Pillet}}]{2001ApJ...547.1148W}
{Westendorp Plaza}, C., {del Toro Iniesta}, J.~C., {Ruiz Cobo}, B., \&
  {Pillet}, V.~M. 2001, \apj, 547, 1148

\end{thebibliography}

\newpage

\begin{appendix}
\section{Removal of scattered light}

\begin{figure*}
\begin{center}
\scalebox{0.45}{\includegraphics{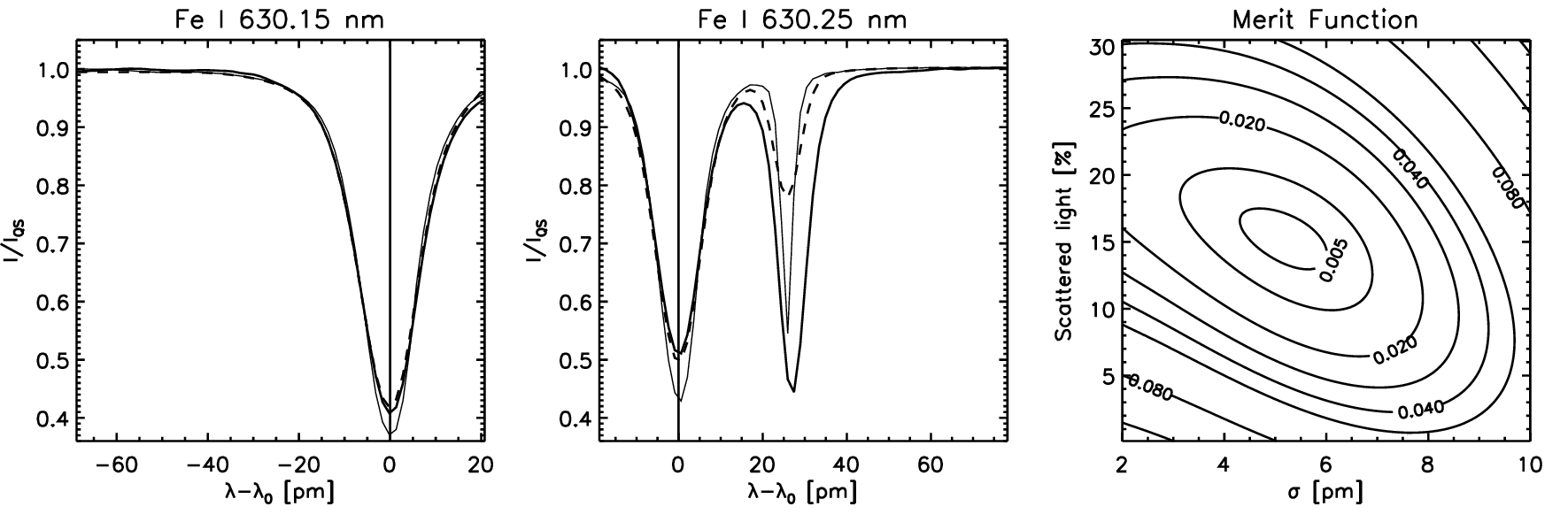}}
\caption{Comparison of the spectra at disk center of \ion{Fe}{i} 630.15~nm
  ({\em left panel}) and \ion{Fe}{i} 630.25~nm ({\em center panel}) as
  observed with FTS (thin solid) and POLIS (thick solid) for the 1 July
  dataset. The dashed lines are the best fits obtained by combining the FTS
  profiles with the fraction of scattered light and the resolving power of
  POLIS as explained in the text. The merit function of the fit is shown in
  the {\em right panel}. Note that the O$_2$ telluric line is not considered
  in the fit.\label{fig:scatter_polis}}
\end{center}
\end{figure*}

\begin{figure*}
\begin{center}
\scalebox{0.67}{\includegraphics{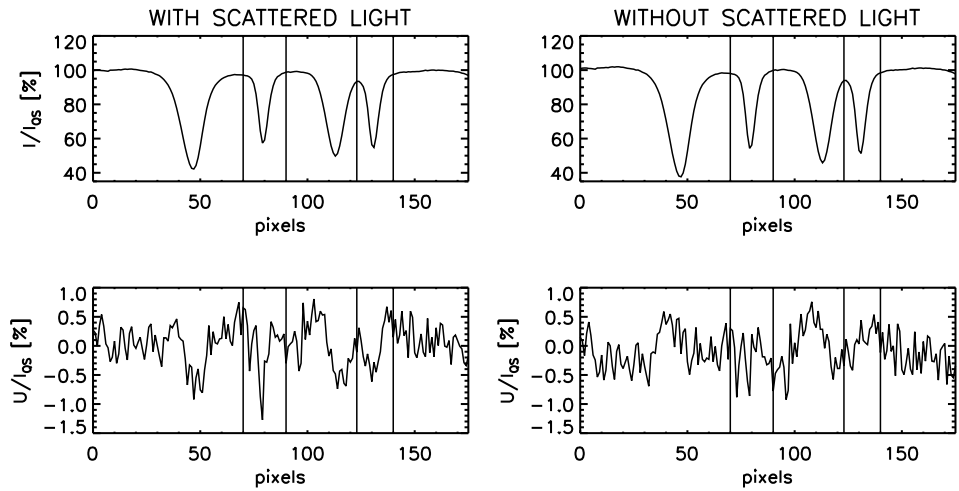}}
\caption{Stokes $I$ ({\em top}) and Stokes $U$ ({\em bottom}) profiles
of the lines at $630$~nm before ({\em left}) and after ({\em right}) correction 
for scattered light. The vertical lines indicate the positions of the 
O$_2$ telluric lines.\label{fig:telluric_polis}}
\end{center}
\end{figure*}

In Fig.~\ref{fig:scatter_polis} we present fits to the disk-center profiles 
of \ion{Fe}{i} 630.15~nm and 630.25~nm recorded on July 1, 2005 using
Eq.~\ref{eq:scatter_light}. As can be seen, the two visible lines are well
reproduced. The merit function of the fit (right panel) shows a minimum for 
$K \simeq 15 \%$ and $\sigma \simeq 6$~pm. The same results are obtained for 
the profiles measured on June 30, 2005. 

Figure~\ref{fig:telluric_polis} shows an example of the effect of the
scattered light correction on the polarization spectra: the original Stokes
$U$ profile displays signatures of the telluric lines, but they disappear
after the correction.

\section{Line parameter maps from the July 1 observations}
\label{sec:appendix}

\begin{figure*}
\begin{center}
\scalebox{.9}{\includegraphics[bb=54 440 435 1091, clip]{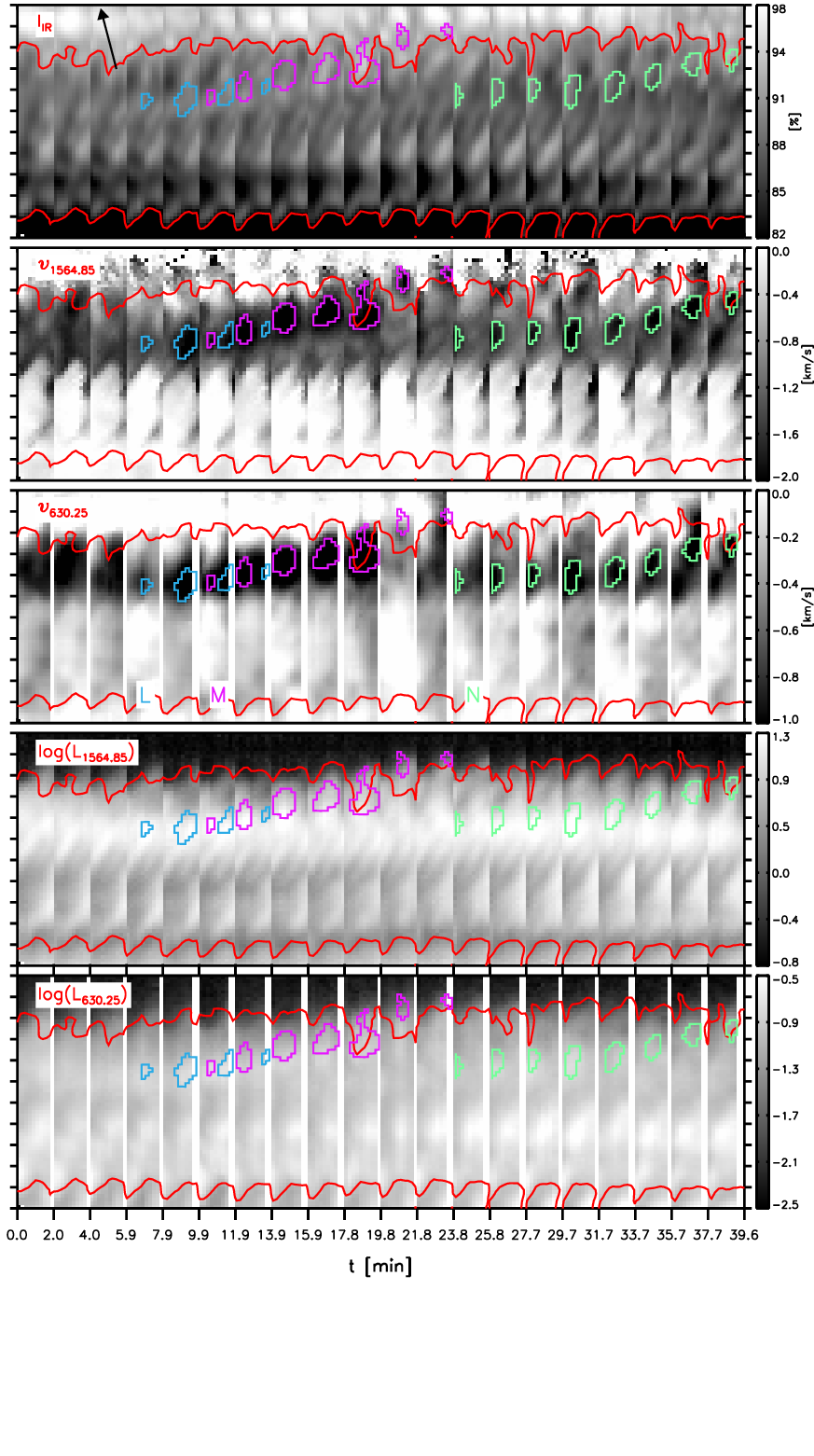}}
\caption{Time evolution of line parameters in the center-side penumbra of AR
10781 on July 1, 2005. From {\em top} to {\em bottom}: Continuum intensity at
1565~nm, Stokes $V$ zero-crossing velocity of \ion{Fe}{i} 1564.85~nm and
\ion{Fe}{i} 630.25~nm, logarithm of the total linear polarization of
\ion{Fe}{i} 1564.85~nm and \ion{Fe}{i} 630.25~nm. Color contours outline the
ECs. The letters at the bottom of the third panel label each EC. Red lines
indicate the inner and outer penumbral boundaries. Each tickmark in the
$y$-axis represent 1\arcsec. The arrow marks the direction to disk
center. \label{fig:01general_1}}
\end{center}
\end{figure*}

\begin{figure*}
\begin{center}
\scalebox{.9}{\includegraphics[bb=54 440 435 1091,clip]{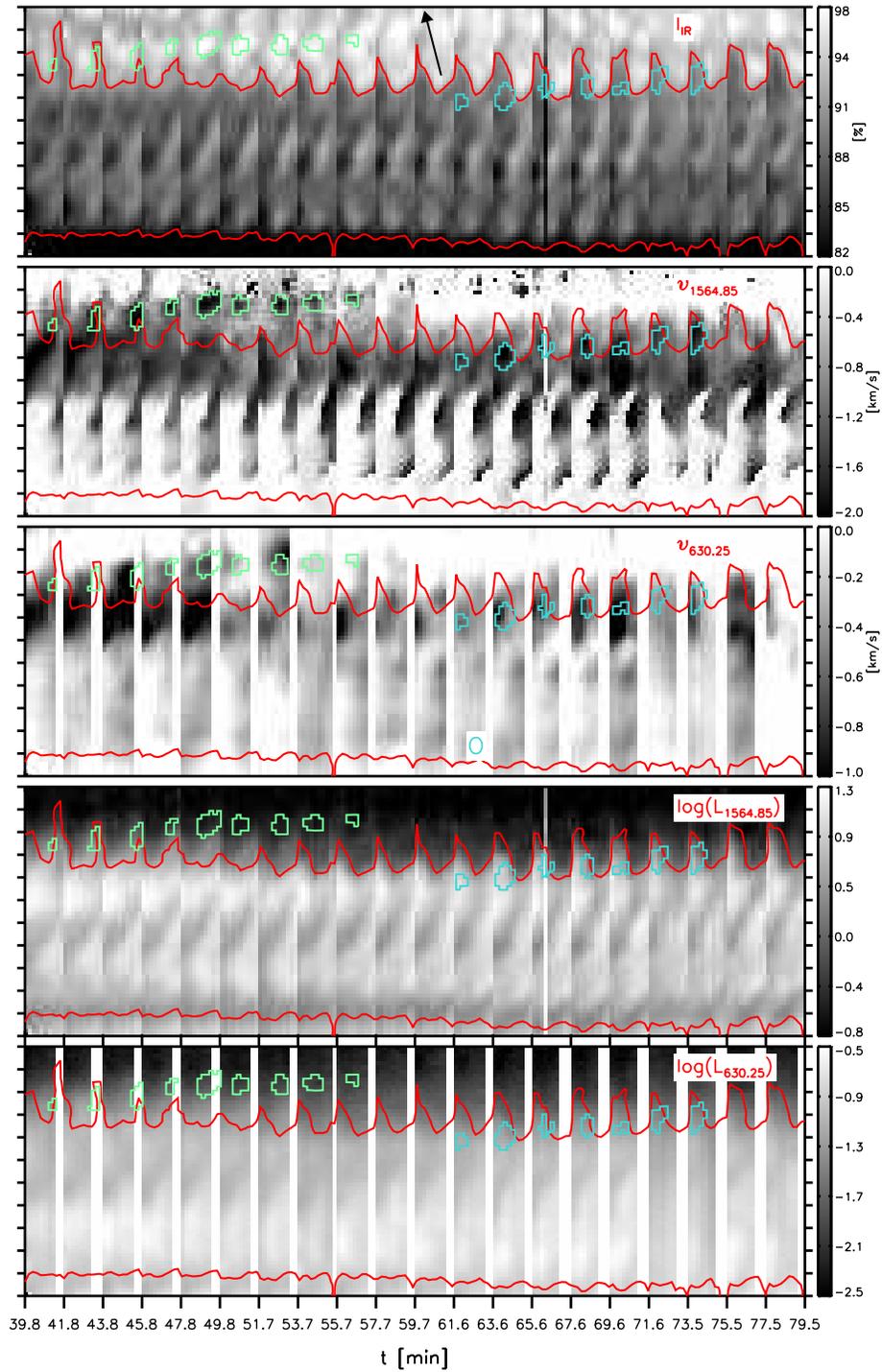}}
\caption{Same as Fig.~\ref{fig:01general_1}, from $t=39.8$ to
$t=79.5$ min.\label{fig:01general_2}}
\end{center}
\end{figure*}

\end{appendix}

\end{document}